\documentclass{jpp}

\usepackage[T1]{fontenc}

\usepackage{setspace}
\usepackage{caption,subcaption}
\usepackage{graphicx}
\usepackage{amsmath, amssymb}
\usepackage{xcolor}
\usepackage{endnotes}
\usepackage{natbib}
\newcommand{\LL} {
\mathcal{L}
}
\newcommand{\MM} {
\mathcal{M}
}

\newcommand{\gm}[2] {
  \mathcal{J}_{#2} #1
}

\newcommand{\del}[1] {
  \delta \! #1
}

\newcommand{\df} {
  \del{f}
}

\newcommand{\dg} {
  g
}

\newcommand{\jj} {
  \ell
}

\newcommand{\kk} {
  m
}

\newcommand{\Gjk} {
  G_{\jj,\kk}
}

\newcommand{\Hev}[1] {
  {\rm He}_{#1}(\nvpar)
}

\newcommand{\Lv}[1] {
  {\rm L}_{#1}(\nmu B)
}

\newcommand{\pderiv}[2] {
  \frac{\partial #1}{\partial #2}
}

\newcommand{\gyavgR}[1] {
  \langle{ #1 \rangle}_{\bf R}
}
\newcommand{\gyavgr}[1] {
  \langle{ #1 \rangle}_{\bf r}
}

\newcommand{\vE} {
  {\bf v}_E
}

\newcommand{\bhat}{
  {\bf \hat{b}}
}

\newcommand{\upar} {
   {u}_\parallel
}

\newcommand{\qpar} {
   {q}_\parallel
}

\newcommand{\qprp} {
   {q}_\perp
}

\newcommand{\nvpar} {
  {v}_\parallel
}

\newcommand{\nmu} {
  \mu
}

\newcommand{\fsa}[1] {
  \langle \langle #1 \rangle \rangle
}

\begin{document}

\title{Laguerre-Hermite Pseudo-Spectral Velocity Formulation of Gyrokinetics}

\author{N.~R.~Mandell\aff{1} \corresp{\email{nmandell@princeton.edu}}, W. Dorland\aff{2,3}, \and M. Landreman\aff{3}}

\affiliation{\aff{1} Princeton Plasma Physics Laboratory, Princeton University, Princeton, NJ 08543, USA
\aff{2} Department of Physics, University of Maryland, College Park,
MD, 20742, USA
\aff{3} Institute for Research in Electronics and Applied Physics,
University of Maryland, College Park, MD, 20742, USA}

\maketitle

\begin{abstract}
First-principles simulations of tokamak turbulence have proven to be
of great value in recent decades. We develop a pseudo-spectral velocity
formulation of the turbulence equations that smoothly interpolates between the highly
efficient but lower resolution 3D gyrofluid representation and the
conventional but more expensive 5D gyrokinetic representation. Our
formulation is a projection of the
nonlinear gyrokinetic equation onto a Laguerre-Hermite velocity-space basis.
We discuss issues related to
collisions, closures, and entropy. 
While any collision operator can be used in the formulation, we highlight a model operator that has a particularly sparse Laguerre-Hermite representation, while satisfying conservation laws and the H theorem.
Free streaming, magnetic drifts, and
nonlinear phase mixing each give rise to closure problems, which we
discuss in relation to the instabilities of interest and to free
energy conservation. We show that the model is capable of
reproducing gyrokinetic results for linear instabilities and zonal flow dynamics. 
Thus the final model is appropriate for the study of
instabilities, turbulence, and transport in a wide range of geometries,
including tokamaks and stellarators. 
\end{abstract}


\section{Introduction}
\label{sec:intro}

Tokamaks and stellarators are magnetic confinement fusion (MCF)
concepts which employ toroidal magnetic fields to confine
thermonuclear plasmas. Steep gradients of density, momentum and
temperature are intrinsic to most MCF concepts. For the most part, the
plasma column itself is macroscopically stable, but these gradients
make it prone to microscopic instabilities and non-thermal
fluctuations. These micro-instabilities are well-described by gyrokinetics \cite[]{Antonsen,Catto78,FChen}.
Many gyrokinetic models, algorithms, and codes have been developed and
deployed over a period of decades. The first simulation algorithms used for gyrokinetics
were particle-in-cell (PIC) algorithms \cite[]{Lee83, DimitsLee, deltaf, Denton, Parker,
  DimitsPRL}. Later, approaches that involved fluid moments
of the gyrokinetic equation were developed. These gyrofluid (or Landau fluid) models
used sophisticated (but linear) closures to model kinetic effects such as Landau damping and finite Larmor
radius (FLR) effects \cite[]{HP,HD,D1,H_sher93,BeerGF,Snyder01}; however, these reduced models 
struggled to accurately predict thermal transport levels, in part due
to inaccurate zonal flow dynamics \cite[]{RH,Cyclone}. The gyrokinetic
codes that became the most widely used emerged around this time
\cite[]{KR,ETG_Dorland00,fsj.ppcf,GYRO}. These Eulerian codes (GS2, GYRO, and
GENE) solve the nonlinear gyrokinetic equations on a fixed,
five-dimensional mesh in phase space $({\bf x}, {\bf v})$. Since the
late 1990's, these codes have been used and cited in hundreds of papers to
analyze small-scale instabilities and turbulence in tokamaks,
stellarators, and laboratory dipole experiments; gyrokinetic codes have also recently been
repurposed for studying turbulence in astrophysical plasmas.\footnote{This includes
analyzing fluctuations in the solar wind and related heliospheric plasmas
\cite[]{AstroGK,Howes06,howes2011gyrokinetic}. Further, gyrokinetic theory has
come to be seen as one of several interesting kinetic limits
associated with larger scale magnetohydrodynamic turbulence
\cite[]{schekochihin2009astrophysical,
  quataert2002magnetorotational}. It is possible that advances in MCF
algorithms could play a role in resolving interesting outstanding
questions in those communities.} There are also many
newer approaches, including semi-Lagrangian \cite[]{GYSELA} and modern
PIC approaches \cite[]{ORB5}. Each approach has distinct advantages, and most codes
have been benchmarked extensively with one another.

A shared weakness of these approaches is the inability to run
reliably with low velocity resolution.  This is despite the fact that
early $\df$ PIC simulations produced accurate estimates of ion heat
flux for as few as 2-4 particles per cell \cite[]{Cyclone}, indicating
that one does not always need to resolve the fine details of the
perturbed distribution function to calculate key fluctuation
properties accurately.  However, using this few particles in $\df$ PIC
is problematic due to Monte Carlo shot noise \cite[]{Nevins05,Lin07} and
inability to resolve highly anisotropic {\it spatial} structures
\cite[]{Idomura2000,Nevins05}.  Further, the basic conservation laws are
not built in, and there is no inexpensive and rigorous way to measure
or control the growth of noise as the simulation progresses. Full-$f$
PIC simulations for tokamaks and stellarators, on the other hand, lack
the flexibility to run with relatively constant precision at low
resolution ({\it i.e.,} with few particles). Conventional Eulerian
gyrokinetic algorithms also lack the flexibility to run with very low
velocity-space resolution while controlling discretization artifacts
associated with velocity-space filamentation, though some codes such
as AstroGK \cite[]{AstroGK} use pseudo-spectral grids to increase the accuracy of
integrals.

In this work we present a pseudo-spectral velocity formulation of
gyrokinetics.  Unlike the Eulerian algorithms described above that
discretize velocity space on a mesh, our approach projects
the gyrokinetic equation onto a velocity basis composed of Laguerre and
Hermite polynomials.\footnote[2]{ The use of Hermite polynomials as a
  velocity basis for Vlasov kinetics has a long history
  \cite[]{grad1949kinetic, armstrong1967numerical, grant1967fourier,
    H_sher93, plunk2014irreversible, parker_dellar_2015,
    kanekar2015fluctuation}. Hermite polynomials have also been used extensively as
  a parallel velocity basis in drift kinetics and gyrokinetics, though
  most approaches have focused on slab geometry \cite[]{Smith97,
    watanabe2004kinetic, hatch2013transition, schekochihin2015phase,
    parker2016}, especially in astrophysical contexts
  \cite[]{zocco2011reduced, loureiro2013fast,
    loureiro2016viriato}. Laguerre polynomials are the natural analog
  for perpendicular velocity space in gyrokinetics, though other
  perpendicular bases have been investigated \cite[]{ParkerThesis,
    landreman2013new}.  } Projecting the distribution function onto
this basis produces fluid-like quantities, which correspond to
density, parallel momentum, {\it etc}. The result is an algorithm
that is equivalent to a generalized gyrofluid system with arbitrarily many moments, 
which at high velocity-space resolution corresponds directly
to conventional gyrokinetics. A key advantage of our approach 
is the flexibility to use very low velocity-space resolution within the same framework,
where the system corresponds precisely to
the well-established gyrofluid models described above. Further,
since our approach produces a system of fluid-quantity conservation
laws, one can maintain fundamental properties like energy
conservation, even at very low resolution.  Any inaccuracy from using
low resolution can be isolated to the shortcomings of closure
assumptions made to allow truncation of the system. Importantly,
however, since the worst case scenario is the level of accuracy of the
original gyrofluid models (which used only 6 moments), our model
should be able to obtain fairly precise estimates of most quantities
of interest at resolutions only slightly higher than that of the
original gyrofluid models. We also present a model collision operator
which has a particularly sparse representation in the Laguerre-Hermite
basis. This collision operator maintains conservation laws and the
H theorem.

The particular application we target is the modeling of MCF turbulence
in the context of whole-device modeling (WDM). The WDM idea is to
simulate the full operation of a tokamak or stellarator, from the
currents in the external magnets to heating systems, to turbulence,
{\it etc.,} with the run-time flexibility to use a range of models for
any given process in any given calculation, depending on the fidelity
required for that process for that simulation run. Our model can thus
run at relatively low velocity resolution as an inexpensive WDM module, 
producing a significant performance advantage over other approaches, 
or at high resolution at costs comparable to the state of the art. 
Thus, a Laguerre-Hermite pseudo-spectral gyrokinetic module for WDM
applications promises high fidelity at modest cost, together with
straightforward validation at high resolution.

This paper is organized as follows: In Section 2, we introduce the gyrokinetic 
equation, which is the starting point of our approach. In Section 3, we derive
our Laguerre-Hermite pseudo-spectral formulation of gyrokinetics. This
includes details of our model collision operator in Section \ref{sec:collisions}.
Section 4 discusses free energy and its conservation in the context of our formulation.
We present linear results of our model in Section 5, and discuss conclusions and future
work in Section 6.

\section{Gyrokinetic description of dynamics} 
\label{sec:gyrokinetic} 

The literature of gyrokinetics and gyrokinetic turbulence is vast and
mature. Our notation and philosophy follow four particular
references: \cite{Antonsen}, \cite{FChen}, \cite{Barnes10}, 
and \cite{Abel13}. In the context of
the latter, this report describes a potentially useful representation
of the nonlinear gyrokinetic equation [Eq.~(108) of \cite[]{Abel13}] in
the absence of strong equilibrium flows and in the absence of
electromagnetic fluctuations.

We begin with a non-dimensional, normalized form of the electrostatic gyrokinetic 
equation:
\begin{align}
\pderiv{g_s}{t} & + \left[v_{ts} v_\parallel \bhat + \gyavgR{\vE} +
                  \frac{\tau_s}{Z_s}{\bf v}_d \right] \cdot \nabla h_s
                  + \gyavgR{\vE} \cdot \nabla F_{Ms}  
                  - v_{ts} \mu \left(\bhat\cdot\nabla B\right) \pderiv{h_s}{v_\parallel} = C(h_s).
\label{gk}
\end{align}
This equation describes the evolution of the fluctuating gyrokinetic
distribution function $g_s=g_s({\bf R}, v_\parallel, \mu, t)$ for a
particular species $s$. All quantities in the above equation
({\it e.g.,} $h,\ g,\ \vE,\ v_\parallel,\ \mu,\ t,\ etc.$) have been
non-dimensionalized, with the specific normalizations explained in
Appendix \ref{app:norm}. The total distribution function,
$F_s = F_{0s} + \delta f_s = F_{Ms} (1 - Z_s \Phi / \tau_s) + h_s$ is a
Maxwellian with a small Boltzmann component
[$\propto \Phi({\bf r}, t)$, the electrostatic potential] and a
general perturbation, $h_s({\bf R}, v_\parallel, \mu, t)$.  The total
{\it gyroaveraged} distribution function,
$\langle F_s \rangle_{\bf R} = F_{0s} + g_s$, with $g_s = h_s - Z_s \gyavgR{\Phi}/\tau_s F_{Ms}$, describes the
probability of finding a particle of species $s$ with guiding center
(or gyrocenter) position ${\bf R}$, velocity parallel to the magnetic
field $v_\parallel$, and magnetic moment $\mu=v_\perp^2/2B$, where
$v_\perp$ is the speed in the plane perpendicular to the magnetic
field. The electrostatic potential
$\Phi=\Phi({\bf r}, t)$ is a function of particle position $\bf r$, where
${\bf r}= {\bf R} + \boldsymbol{\rho}$ so that $\boldsymbol{\rho}$ is
the gyroradius vector that rotates at the gyrofrequency $\Omega$ and
points from the gyrocenter (at $\bf R$) to the particle (at $\bf
r$). The equilibrium magnetic field ${\bf B}$ has magnitude
$B=B({\bf r})$ and direction $\hat{\bf b} = {\bf B}/B$.

The equilibrium distribution function is a Maxwellian (which we take to have no flows),
\begin{equation}
F_{0s} = F_{Ms} = 
\frac{n_s}{(2\pi v_{ts}^2)^{3/2}}e^{-v_\parallel^2/2- \mu B},
\end{equation}
and the fluctuating gyroaveraged distribution function satisfies
$g\ll F_M$. We also have the following dimensionless species
parameters: equilibrium density $n_s$, equilibrium temperature
$\tau_s$, mass $m_s$, charge $Z_s$, and thermal velocity
$v_{ts}=\sqrt{\tau_s/m_s}$.

The notation $\langle \dots \rangle_{\bf R}$ denotes a gyroaverage
taken at constant ${\bf R}.$\footnote{In Fourier space, the gyroaveraging operation
(whether of a function of $\bf r$ at constant $\bf R$, or vice versa) 
is simply a multiplication by the Bessel function
  $J_{0,s}=J_0(\sqrt{2\mu B b_s})$, where $b_s = k_\perp^2
  \rho_s^2$. Thus we will use the notation
  $\gyavgR{A({\bf r})}=\gyavgr{A({\bf R})}=J_0 A$ interchangeably, with the
  understanding that $J_0 A$ is a Fourier space operation.}  The
gyroaveraged $\bf E\times B$ velocity is
$\gyavgR{\vE} = \bhat\times \nabla \gyavgR{\Phi} $, and the magnetic
drift velocity is
${\bf v}_d = (v_\parallel^2 + \mu B) \bhat\times\nabla B/B^2$. The
collision operator, which we define in Section \ref{sec:collisions},
is denoted by $C(h)$. Finally, the
potential is determined by the quasineutrality equation, which we also
write in terms of $h$:
\begin{equation}
\sum_s Z_s n_s \int d^3{\bf v}\ \gyavgr{h_s}= \sum_s \frac{n_s Z_s^2}{\tau_s} \Phi. \label{qneuts}
\end{equation}
In the limit of a single hydrogenic ion species with Boltzmann electrons, this reduces to 
\begin{equation}
\int d^3{\bf v} \, \gyavgr{h} = 
  \tau_e^{-1} \, \left[\Phi -\langle \langle \Phi \rangle \rangle\right] + \Phi, 
\label{qneut}
\end{equation}
where $\tau_e = T_e/T_i$. (Note that several authors choose the
reciprocal definition for $\tau_e$.) The flux-surface average is
denoted by $\fsa{\dots}$. This term comes from the standard correction
for Boltzmann electrons \cite[]{D1}, which ensures that the
flux-surface-averaged electron density perturbation vanishes.

Eqs.~(\ref{gk}) and
(\ref{qneut}) describe self-consistent electrostatic gyrokinetic
dynamics in field-line-following coordinates, including magnetically
trapped particles and as yet unspecified collisional physics. 
These equations can be solved in general geometry
(tokamak and stellarator) to find the instabilities, fluctuation
spectra, and turbulent fluxes of particles and energy. Landau damping,
trapped particle effects, and other ``kinetic'' phenomena are
described by this comprehensive approach.

\section{Laguerre-Hermite pseudo-spectral formulation}
\label{section:HL}

Our basic approach is to project the velocity dependence of the 
fluctuating distribution functions $h$ and $g$ onto orthonormal polynomials. 
The result is a spectral representation where the
amplitudes of the polynomial basis functions, which can be interpreted
as fluid moments, become the
dynamical variables. Thus instead of keeping
track of $g(v_\parallel, \mu B)$ with a finite-difference algorithm on
a discretized $(v_\parallel, \mu B)$ grid, we evolve coupled
fluid-like quantities, such as density, parallel momentum, {\it etc.},
which emerge from the projection. 
Specifically, we project onto Hermite polynomials in the $v_\parallel$
coordinate and Laguerre polynomials in the $\mu B$ coordinate.
We choose this basis in part because the resulting system is identical at low resolution to
the gyrofluid models of \cite{D1}, and \cite{BeerGF}. The
Laguerre and Hermite
polynomials are also orthogonal with respect to a Maxwellian.
Further, the Laguerre and Hermite polynomials are eigenfunctions of our model collision operator, 
described in Section \ref{sec:collisions}.

\subsection{Pseudo-spectral in phase space}

Our approach is perhaps best
understood by analogy with pseudo-spectral Fourier methods. 
Consider a Fourier decomposition for the
spatial coordinates in the gyrokinetic equation.  By using
field-line-following coordinates to avoid irregular domains
\cite[]{BCH}, Fourier techniques can be readily applied; periodic (and
otherwise appropriate) domains are guaranteed by the asymptotics of
the problem \cite[]{Abel13}.  Such a Fourier decomposition has the
advantage of allowing spectrally accurate evaluation of the
derivatives because the spatial derivative has Fourier harmonics as
its eigenfunctions [$d/dx \exp{(ikx)} = ik \exp{(ikx)}$].  Conversely,
achieving spectral accuracy in evaluating derivatives with a
finite-difference, real-space representation on an $n$-point grid
requires $n$-point stencils, which is expensive.  There are costs, of
course, that reduce the advantages of a Fourier
representation. Nonlinearity, in particular, necessitates the
evaluation of convolutions in a strict Fourier basis: quadratic
nonlinearities introduce spectral convolutions with $O(n^2)$ terms.
Pseudo-spectral algorithms reduce this operator count to
$O(n \log{n})$ by using fast discrete transforms to evaluate the
nonlinear terms in real space, without the loss of spectral accuracy.

Our new approach takes this a step further by using pseudo-spectral
methods not only in configuration space but also in velocity space.
Instead of working with $f({\bf R}, {\bf v})$, we Fourier
transform in the spatial coordinate $\bf R$ and Laguerre-Hermite transform in the velocity coordinate $\bf v$ to
develop spectral equations governing the evolution of
$f({\bf k}, {\bf p})$. Here, $\bf k$ is the usual Fourier wavenumber,
and ${\bf p}=(\ell,m)$ is the ``wavenumber'' in the Laguerre-Hermite dual to
velocity space.
The advantages of the Fourier decomposition discussed in the previous
paragraph apply to velocity space as well. Just as a Fourier
decomposition in space has the advantage of spectrally accurate
evaluation of spatial derivatives, our Laguerre-Hermite spectral
velocity decomposition allows spectrally accurate evaluation of the
velocity derivatives, which appear in Eq. (\ref{gk}) and in the collision operator. 
Nonlinearities may be
evaluated pseudo-spectrally relatively inexpensively in
$({\bf R}, {\bf v})$ space with the use of efficient transforms and
appropriately chosen grids. We expect this pseudo-spectral formulation
of gyrokinetics to be a useful addition to the community toolbox.

There are disadvantages to our approach, too. In addition to
derivatives in velocity space, there are multiplications by velocity
in the gyrokinetic equation. The canonical example is the free
streaming of particles along the magnetic field
$(\sim v_\parallel d/dz)$, which introduces a term like
$i k_z v_\parallel$ to the fluctuation equations which describe a
general fluctuation $\df\, ({\bf k}, {\bf v})$. In $({\bf R, v})$
space, a term like this can be evaluated with a finite-difference
scheme \cite[]{GYRO,ETG_fsj}. In $({\bf k, v})$ space, this can be
evaluated with an integrating factor.  In our $({\bf k, p})$ space
formulation, this term couples the equations describing the evolution
of the various Hermite polynomial moments of
$\df\,(v_\parallel)$. This is the mathematical manifestation of the
physics of phase mixing, and it introduces the need to close the
finite set of Hermite moments used.  At high enough resolution,
collisions (even if weak) provide physical closure, but we anticipate
the key benefit of using a variable spectral velocity representation
will be in the low resolution limit. The model can be regarded as a
generalized gyrofluid system, with the flexibility to increase the 
number of moments to achieve any desired level of
accuracy. {In the limit of a large number of
  moments, the model is equivalent to traditional grid-based
  gyrokinetic algorithms.} 

\subsection{Laguerre \& Hermite functions}
We begin by defining expansion and projection functions in our basis. 
The Laguerre expansion and projection functions are given (respectively) by
\begin{equation}
\psi^\ell (\mu B) = (-1)^\ell e^{-\mu B} \Lv{\ell},
\qquad\qquad 
\psi_\ell (\mu B) = (-1)^\ell \Lv{\ell},
\label{eq:Lbases}
\end{equation}
where
\begin{equation}
{\rm L}_\ell(x)=\frac{e^x}{\ell!}\frac{d^\ell}{dx^\ell} x^\ell e^{-x}
\end{equation}
are the Laguerre polynomials. 
The Hermite expansion and projection functions are given (respectively) by
\begin{equation}
\phi^m (v_\parallel) = \frac{e^{-v_\parallel^2/2} \Hev{m}}{\sqrt{(2\pi)^3 m!}},
\qquad\qquad 
\phi_m (v_\parallel) = \frac{\Hev{m}}{\sqrt{m!}},
\label{eq:Hbases}
\end{equation}
where
\begin{equation}
{\rm He}_m(x)=(-1)^m e^{x^2/2}\frac{d^m}{dx^m}e^{-x^2/2}
\end{equation}
are the probabilists' Hermite polynomials.\footnote{Note that if one instead defined
$v_t=\sqrt{2 T_{0}/m}$, as is chosen by many authors, one would use the physicists' Hermite
polynomials, ${\rm H}_m(x)=(-1)^m e^{x^2} 
\frac{d^m}{dx^m}e^{-x^2}$. This would result in many coefficients in the following equations being different by factors of $\sqrt{2}$.
}

These definitions yield orthonormality conditions, 
\begin{gather}
\int_{0}^\infty d \mu B\ \psi^k(\mu B) \psi_\ell(\mu B)  = \delta_{k \ell} 
\qquad \qquad
2\pi\int_{-\infty}^\infty d v_\parallel\ \phi^m(v_\parallel)
\phi_n(v_\parallel)  = \delta_{m n},
\label{orthonormality}
\end{gather}
where $\delta_{ij}$ is the Kronecker delta. 
The expansion functions satisfy recurrence relations given by
\begin{gather}
\mu B \psi^\ell(\mu B) = (\ell+1) \psi^{\ell+1}(\mu B) + (2\ell+1) \psi^\ell(\mu B) +
\ell \psi^{\ell-1}(\mu B),
\label{recurL}\\
v_\parallel \phi^m(v_\parallel) =
\sqrt{m+1}\phi^{m+1}(v_\parallel) + \sqrt{m}
\phi^{m-1}(v_\parallel).
\label{recurHe} 
\end{gather}
We will also need the following derivative relations:
\begin{equation}
B\pderiv{\psi^\ell}{B} = -(\ell+1) \, \left(\psi^{\ell+1} + \psi^\ell \right),
\qquad\qquad
\pderiv{\phi^m}{v_\parallel} = -\sqrt{m+1} \, \phi^{m+1}. 
\label{basisDerivatives}
\end{equation}
Recall that all velocities here are normalized to the species thermal velocity. 
In the following sections, we suppress the polynomial arguments for concision. 

\subsection{Laguerre-Hermite expansion}

The Laguerre and Hermite expansion functions defined above 
form a complete set. One can therefore expand the perturbed distribution
functions $g$ and $h$ as
\begin{equation}
g = \sum\limits_{\ell=0}^\infty \sum\limits_{m=0}^\infty
\psi^\ell \phi^m \Gjk, \qquad \qquad h = \sum\limits_{\ell=0}^\infty \sum\limits_{m=0}^\infty
\psi^\ell \phi^m H_{\ell,m}, \qquad  \label{dg}
\end{equation}
with spectral amplitudes defined by the projection of $g$ and $h$ onto
the Laguerre-Hermite basis, given by
\begin{gather}
\Gjk = 2\pi \int_{-\infty}^\infty d v_\parallel
\int_0^\infty d\mu B\ \psi_\ell \phi_m \dg, \label{eq:glm} \\
H_{\ell,m} = 2\pi \int_{-\infty}^\infty d v_\parallel
\int_0^\infty d\mu B\ \psi_\ell \phi_m h =  G_{\ell,m} + \frac{Z_s}{\tau_s} \mathcal{J}_\ell \Phi \delta_{m 0}.
\end{gather}
Note that $G_{\ell,m} = H_{\ell,m}$ for $m\neq0$. Physically, the $\Gjk$ ($H_{\ell, m}$) are orthonormal fluid moments of $\dg$ ($h$).  Each $G_{\ell, m} = G_{\ell, m} ({\bf R}, t)$ is a guiding center function of space and time. We will frequently
work with the perpendicular Fourier components of $G_{\ell, m} ({\bf
k}_\perp, z, t)$ without changing notation when the context is clear. 

We also
note that there is a direct correspondence with the gyrofluid moments defined by \cite{BeerGF},
\begin{align}
\left( G_{0,0},\ G_{0,1},\ \sqrt{2} G_{0,2} ,\ \sqrt{6} G_{0,3},\ G_{1,0},\ G_{1,1} \right) =
  \left(n,\ u_\parallel,\ T_\parallel,\ q_\parallel,\ T_\perp,\ q_\perp
  \right), \label{Beermoments}
\end{align}
where each moment on the right appears exactly as defined by Beer,
with the understanding that the normalizations have all been
applied. Thus our choice of the Laguerre-Hermite basis allows us
to extend the Beer model to arbitrary number of moments.

Working from Eq.~(\ref{dg}), we can expand the following derivatives that appear in the gyrokinetic
equation:
\begin{gather}
  \bhat \cdot \nabla h =
  \sum_{\ell,m=0}^\infty \psi^\ell \phi^m \bhat \cdot \nabla H_{\ell,m}
  - (\ell+1) \left( \psi^{\ell+1} + \psi^\ell\right) \phi^m  H_{\ell,m} \, \bhat \cdot \nabla \ln B, \label{gradpar}\\
  \bhat \times \nabla h =
  \sum_{\ell,m=0}^\infty \psi^\ell \phi^m \, \bhat \times \nabla \, H_{\ell,m} \label{gradperp}, \\
  \pderiv{h}{v_\parallel} =
  - \sum_{\ell,m=0}^\infty \sqrt{m+1} \, \psi^\ell \phi^{m +1} \, H_{\ell,m}.
\end{gather}

Electric fields are everywhere gyroaveraged in the gyrokinetic equation,
with the gyroaveraging operation expressed in Fourier space as multiplication by the Bessel function
$J_0$. We can conveniently expand $J_0$ in terms of our Laguerre functions as
\begin{equation}
J_0(\sqrt{2\mu B b}) = \sum_{\ell=0}^\infty \psi_\ell \frac{1}{\ell!} \left(-\frac{b}{2}\right)^\ell e^{-b/2}
 \equiv \sum_{\ell=0}^\infty \psi_\ell \mathcal{J}_\ell (b), \label{j0HL}
\end{equation}
where $b=b_s=k_\perp^2 \rho_s^2$, and we have defined 
\begin{equation}
\mathcal{J}_\ell \equiv \frac{1}{\ell!} \left(-\frac{b}{2}\right)^\ell e^{-b/2} \label{jflr}.
\end{equation}
This can be interpreted as the amplitude of the gyroaveraging operator
in Fourier-Laguerre space.  This expression is consistent with the
physical picture that gyroaveraging greatly attenuates short
wavelengths $(b > 1)$ but has little effect at long wavelength $(b<1)$
when $\ell=0$. Gyroaveraging results in quite a strong attenuation at
long wavelength for $\ell>0$, since $\gm{}{\ell} \sim b^\ell$ for small
$\ell$. Finally, the $1/\ell!$ dependence of $\mathcal{J}_\ell$ means that the
contribution of high-$\ell$ Laguerre moments to the gyroaveraged
potential is small. 

Note that the $e^{-b/2}$ in the gyroaveraging operator makes our basic
FLR approach consistent with that of \cite{Brizard}. Dorland made alternative
FLR approximations in his gyrofluid model in order to obtain better accuracy at low
Laguerre resolution \cite[]{D1}. Details on the FLR accuracy of our model are 
presented in Appendix \ref{appendix:flr}. Nonlinearly, FLR effects give rise to nonlinear phase-mixing,
which was modeled in \cite[]{D1} but not in \cite[]{Brizard}. 

\subsection{Laguerre-Hermite fluid equations}

We now derive the infinite set of coupled 3D fluid equations
governing the evolution of the Laguerre-Hermite spectral amplitudes
$\Gjk$. The procedure is straightforward: we work term by term in the
gyrokinetic equation, first using the identities above to expand each
term as an infinite Laguerre-Hermite sum, and then using the
orthogonality relations to project the terms onto the Laguerre-Hermite
basis. To illustrate this procedure, we show as an example how we project 
the magnetic drift term in the gyrokinetic equation in Appendix \ref{app:toroidal}.

This effectively reduces (without approximation) the 5D gyrokinetic
equation to an infinite
collection of coupled 3D fluid equations [Eq.~(\ref{glmevolve}), below]. Coupling arises from
finite Larmor radius (FLR) effects, the ${\bf E}\times{\bf B}$ and
toroidal drifts, collisions, and flows along field lines, including
flows associated with magnetic trapping. 

The projection of the nonlinear term in Eq.~(\ref{gk}) involves the
convective derivative,
\begin{align}
{d G_{\ell,m} \over dt} &\equiv \pderiv{G_{\ell,m}}{t}   
+ \sum\limits_{k=0}^\infty\ \sum_{n=|k-\ell|}^{k+\ell}
\alpha_{k \ell n} \, (\gm{\vE}{n}) \cdot \nabla G_{k,m} \notag\\
&= \pderiv{G_{\ell,m}}{t}   
+ \sum\limits_{k=0}^\infty\ \sum_{n=|k-\ell|}^{k+\ell}
\alpha_{k \ell n} \, (\gm{\vE}{n}) \cdot \nabla H_{k,m},
\label{convective}
\end{align}
where the equivalence of the expressions with $G$ and $H$ follows from the fact that
$\langle \vE \rangle \cdot \nabla h = \langle \vE \rangle
\cdot \nabla g$ since $\langle \vE \rangle \cdot \nabla \langle \Phi \rangle = 0$. 
The convolution in Eq.~(\ref{convective}) arises from finite Larmor radius (FLR)-induced coupling, and in
particular accounts for nonlinear FLR phase mixing (NLPM) \cite[]{D1}. 
The convolution coefficients are given by
\begin{align}
\alpha_{k \ell n} = \int_0^\infty d \mu B\ \psi_k \psi^\ell \psi_n = \sum_{j}
\frac{(k + \ell -  j)! \quad 2^{2j-k-\ell+n}}{(k-j)!(\ell-j)!(2j-k-\ell+n)!(k+\ell-n-j)!},
\label{Ckmn}
\end{align}
as first calculated by \cite{Watson38}, where the
summation limits are set by the requirement that all factorials have
non-negative arguments. This requirement is consistent with the
bandwidth limits in the sum over $n$ in (\ref{convective}).  

Note that instead of evaluating this term as a convolution, one could use a 
pseudo-spectral approach by transforming to
$(v_\parallel, \mu B)$ coordinates and evaluating the nonlinearity as
a pointwise multiplication. This pseudo-spectral alternative, which is
described in Appendix \ref{app:HLeval}, requires efficient implementations of
the transforms described by Eqs.~(\ref{dg}) and (\ref{eq:glm}).  These transforms can be evaluated 
via matrix multiplication, which is efficient unless one uses quite a
large number of Laguerre moments.\footnote{
 In part, this is because the
required matrix elements can be precomputed, and also
because modern processors (both GPU and CPU) have extremely efficient
implementations of matrix-vector multiplication.}

In terms of Laguerre-Hermite basis functions, Eq.~(\ref{gk}) is thus written
\begin{align}
{d G_{\ell,m} \over dt} & 
  + v_{ts} \nabla_\parallel \left( \sqrt{m+1} \, H_{\ell,m+1} + \sqrt{m} \, H_{\ell,m-1} \right) 
  \notag \\ & 
  +v_{ts}  \Big[ - (\ell+1) \, \sqrt{m+1} \, H_{\ell,m+1} - \ell \, \sqrt{m+1} \, H_{\ell-1,m+1} \notag
  \\
  & \qquad\qquad+ \ell \, \sqrt{m} \, H_{\ell,m-1} 
  + (\ell+1) \, \sqrt{m} \, H_{\ell+1,m-1} \Big] \nabla_\parallel \ln B
  \notag \\ & 
   + i \omega_{ds} \Big[ \sqrt{(m+1)(m+2)} \, H_{\ell,m+2}
   + (\ell+1) \, H_{\ell+1,m} \notag \\
    &\qquad\qquad+ 2 \, (\ell+m+1) \, H_{\ell,m} 
   + \sqrt{m (m-1)} \, H_{\ell,m-2} + \ell \, H_{\ell-1,m} \Big]
  \notag \\ & 
   = D_{\ell,m} + C_{\ell,m}.
\label{glmevolve}
\end{align}
Parallel convection, including bounce motion induced by magnetic
trapping in the equilibrium magnetic field, is described by the terms
proportional to $\nabla_\parallel \equiv \bhat \cdot \nabla$. Toroidicity
gives rise to the terms proportional to $i \omega_{ds}= i \omega_d
(\tau_s/Z_s)$, where $i \omega_{d} \equiv (1/B^2){\bhat}\times\nabla
B\cdot\nabla.$ Drive terms from equilibrium gradients, denoted by $D_{\ell,m}$, are given by
\begin{align}
D_{\ell,m=0} & =  i \omega_* \, \left[\frac{1}{L_{ns}} \gm{\Phi}{\ell} +
            \frac{1}{L_{Ts}}  \left[ \ell \gm{\Phi}{\ell-1} 
            + 2\ell \gm{\Phi}{\ell} + (\ell+1) \gm{\Phi}{\ell+1}
               \right] \right]
 \notag \\ 
D_{\ell,m=2} & =  \frac{1}{\sqrt{2}} i\omega_* \, \frac{1}{L_{Ts}} \,
               \gm{\Phi}{\ell} \notag
 \\
D_{\ell, m} & = 0 \qquad {\rm otherwise},
\end{align}
where $i \omega_*\equiv -\nabla \Psi \cdot
\bhat\times\nabla$, and $L_{ns}$ and $L_{Ts}$ are the normalized density and temperature
gradient scale lengths, respectively.
The collision terms, denoted by $C_{\ell,m}$, are presented in Section \ref{sec:collisions}. 

To complete the Laguerre-Hermite equation set, we must use
Eq.~(\ref{qneuts}) to find the potential $\Phi$, given $h$.
The left-hand side of Eq.~(\ref{qneuts}) involves the non-Boltzmann
part of the particle space density $\bar{n}=\bar{n}({\bf r})$, which is given by 
\begin{equation}
\bar{n} = \int d^3{\bf v} \langle h \rangle_{\bf r} = \int d^3{\bf v} J_0 h = \sum_{\ell=0}^\infty 
 \gm{H_{\ell,0}}{\ell},
\label{eq:nbar}
\end{equation}
where we have used $\gyavgr{h}$ and $J_0 h$ interchangeably for the gyroaverage of $h$, with the understanding that in the latter case we have taken the Fourier transform so that $h=h({\bf k}_\perp)$.
For hydrogenic plasma with Boltzmann electrons, the quasineutrality equation reduces to Eq. (\ref{qneut}), which projects to 
\begin{equation}
\sum_{\ell=0}^\infty \gm{H_{\ell,0}}{\ell} = 
  \tau_e^{-1} \, \left[\Phi -\langle \langle \Phi \rangle \rangle\right] + 
   \Phi. \label{hlmqneut}
\end{equation}
Equivalently, quasineutrality can be expressed in terms of $G_{\ell,0}$,
as 
\begin{equation}
\sum_{\ell=0}^\infty \gm{G_{\ell,0}}{\ell} = 
  \tau^{-1} \, \left[\Phi -\langle \langle \Phi \rangle \rangle\right] + 
  \left[ 1 - \sum_{\ell=0}^\infty \mathcal{J}_\ell^2 \right] \Phi. \label{glmqneut}
\end{equation}
When we truncate the system at some maximal Laguerre moment
$\mathcal{L}$, we truncate the sum of the left hand side of
Eq.~(\ref{hlmqneut}) at $\mathcal{L}$. In Eq.~(\ref{glmqneut}), this
is equivalent to truncating the sums on {\it both} the left and right
hand side of Eq.~(\ref{glmqneut}) at $\mathcal{L}$. This maintains
consistency in the FLR accuracy of the left and right hand side, which
is most clear when we express quasineutrality in terms of
$H$. Further, note that
$\sum_{\ell=0}^\infty \mathcal{J}_\ell^2 = \Gamma_0 = \int d^3{\bf v} J_0^2 F_M$,
where $\Gamma_0=\Gamma_0(b) = I_0(b) e^{-b}$, and
$I_0(b) = J_0(ib)$ is the modified Bessel function.

\subsection{Collision operator} \label{sec:collisions}
To model collisional physics, we generalize the Dougherty
 collision operator \cite[]{Dougherty}, which is itself a generalization
of the Lenard-Bernstein collision operator \cite[]{LB}. Though it is a
simplified, model collision operator, the Dougherty operator has
several appealing properties. It describes pitch angle scattering,
energy diffusion and slowing down. It vanishes on (and only on) a Maxwellian, and it
conserves number, momentum and energy. Thus it satisfies the
appropriate H-theorem, and drives the plasma to thermal equilibrium in
the long-time limit. These properties were recognized and emphasized
by \cite{Anderson}. Below, we gyroaverage the
Dougherty operator and express the result in normalized
$(v_\parallel, \mu)$ coordinates. The result is a gyrokinetic operator
that can be compactly expressed in terms of Laguerre-Hermite polynomials,
which are eigenfunctions of the differential part of the operator, while preserving all of
the above properties.
\subsubsection{Definition}

The starting point is 
\begin{equation}
C(f) = \nu \, {\partial \over \partial {\bf v}} \cdot \left[ 
{\bf D}[f]({\bf v}) \cdot {\partial f \over \partial {\bf v} }
+ P[f]({\bf v}) f
\right],
\label{eq:FP}
\end{equation}
where ${\bf D}$ is a velocity space diffusion tensor that is a
functional of $f$, and $P$ is a field-particle operator, also a
functional of $f$ in the general case.  The collision frequency $\nu$
is constant, {\it i.e.,} independent of velocity. Dougherty
approximated ${\bf D}$ by $T/M$, where $M$ is the particle mass and $T$ is the temperature, allowing $T = T_0 + \delta T$. For
$P[f]$, he used ${\bf v}-{\bf u}$, with the flow ${\bf u} = {\bf u}_0
+ \delta{\bf u}$. Below, we make the further approximation that
${\bf u}_0 = 0$, though this could be relaxed in the future if
desired.

With these assumptions, we can then linearize to obtain
\begin{equation}
 C(\df) = \nu {\partial \over \partial {\bf v}} \cdot \left[ {T_0
    \over M} {\partial \df \over \partial {\bf v}} + {\delta T \over
    M} {\partial F_M \over \partial {\bf v}} + {\bf v} \, \df - \delta {\bf
    u} \, F_M \right].
\label{eq:FP2}
\end{equation}
By construction, $C(F_M) = 0$. Number is automatically conserved because
$C$ has the form of a velocity-space divergence. Momentum and
energy are conserved by the field-particle terms, which are $(\propto
\delta T, \delta {\bf u})$. The forms of the restoring terms are explicitly constructed 
to preserve the properties listed above. 

In gyrokinetic variables, recall from above that the perturbed distribution function can be
written as $
\df = h - {Z_s / \tau_s} \Phi F_M.
$ The velocity space structure of the Boltzmann component $(\propto
\Phi)$ is Maxwellian, and thus
$
C(\df) = C(h). 
$
Note that in general $C(h) \neq C(g)$ due to the velocity dependence of the gyroaveraging operator.

We gyroaverage this collision operator in the standard
way \cite[]{CattoTsang}. Collisions occur at fixed position ${\bf r}$ rather than at fixed
gyrocenter, ${\bf R}$. Thus, the collision operator must be
evaluated at fixed ${\bf r}$. Only at the end does one transform back
to guiding center position ${\bf R}$, as required by the fact that
Eq.~(\ref{gk}) describes the evolution of $g({\bf R})$.

Keeping track of these dependences is slightly awkward, because $h =
h({\bf R})$. Following \cite{Abel08} we express
the distribution function $h$ in ${\bf k}$-space to facilitate the
required manipulations:
\begin{align}
\Big\langle{ C_{\bf r}(h)}\Big\rangle_{\bf R} = &
\Bigg\langle {C \left(\sum_{\bf k} e^{i {\bf k}\cdot{\bf R}} h_k\right)} \Bigg\rangle_{\bf R}
= \sum_{\bf k} \Big\langle e^{i {\bf k}\cdot{\bf r}}
C \left( e^{-i {\bf k}\cdot{\bf \rho}} h_k\right)\Big\rangle_{\bf R}
  \notag \\
= & \sum_{\bf k} e^{i {\bf k}\cdot{\bf R}} 
 \Big\langle{ e^{i {\bf k}\cdot{\bf \rho}} C \left( e^{-i {\bf
    k}\cdot{\bf \rho}} h_k\right)}\Big\rangle_{\bf R}
\equiv C(h).
\label{cphase}
\end{align}
In this form, it is less difficult to perform the gyroaverages.  The
phase factors combine to describe classical diffusion
$\propto -\nu \, (k_\perp^2 \rho^2) \, h$. (Number, momentum and
energy are conserved before and after gyroaveraging, as discussed
below and in Appendix \ref{app:collisions}.)  Setting aside the field-particle
(conservation) terms, in cylindrical $(v_\parallel, \mu)$ coordinates
the result of gyroaveraging the velocity space derivative (or test
particle diffusion) part of the collision operator is
\begin{equation}
C(h) 
= \nu \left[ {\partial \over
     \partial v_\parallel} \left({\partial \over \partial v_\parallel}
   + v_\parallel \right) 
 + 2 {\partial \over \partial \mu} \left( {\mu \over B}
     {\partial \over \partial \mu} + \mu \right) - k_\perp^2 \rho^2
     \right] h.
\label{eqCh}
\end{equation}
Pitch-angle scattering is a particularly important dynamical process
in many experimental scenarios. Our
collision operator, even as it is expressed in cylindrical coordinates
[Eq.~(\ref{eqCh})], describes pitch-angle scattering. Parenthetically, note that
one can make pitch-angle and energy scattering manifest by changing coordinates to $(v, \xi)$, with $\xi = v_\parallel/v$:

\begin{equation}
 C(h)
= \nu \left\{ \frac{1}{v^2}
{\partial \over  \partial \xi} \left[ \left( 1 - \xi^2\right) 
{\partial \over \partial \xi} \right]
 + \frac{1}{v^2} {\partial \over \partial v} \left[ v^2
     {\partial  \over \partial v} + v^3 \right] - k_\perp^2 \rho^2
     \right\} h.
\end{equation}

The field-particle terms are calculated\footnote{We note in passing
  that there would be no difference if one instead integrated
  $\delta f$ at fixed ${\bf r}$ to find these quantities. There are no
  flows in the equilibrium by assumption, and $\int (v_\parallel^2-1)
  F_M = \int (\mu B - 1) F_M = 0$.}
  by
integrals of $h$ at fixed ${\bf r}$. The full expression required for
the momentum conserving terms in the collision operator is
\begin{equation}
 \frac{\delta {\bf u}\cdot {\bf v}}{v_{ts}^2} =  {\bf \bar{u}} \cdot {\bf v} =
  \left[ \bar{u}_\parallel  v_\parallel J_0 + 
    \bar{u}_\perp v_\perp J_1 \right].
\label{coll:allmom}
\end{equation}
Here, the perturbed parallel flow of
particles (not guiding centers) is
\begin{equation}
 \bar{u}_\parallel \equiv \int d^3{\bf v} \, J_0\, v_\parallel \, h.
\label{coll:mom}
\end{equation}
The standard factor of $J_0$ in the integrand arises from integrating
at fixed ${\bf r}$, as dictated by the spatial locality of
collisions, which requires a gyroaverage of $h({\bf R})$. The integral for perpendicular momentum conservation is
also standard, involving
\begin{equation}
 \bar{u}_\perp \equiv \int d^3{\bf v} \, J_1\, v_\perp \, h, 
\label{coll:perp}
\end{equation}
where $J_1(a) = -d/da \, J_0(a)$.  Without the field-particle terms, the collision operator also fails to
conserve energy. The total perturbed energy consists of both parallel
and perpendicular energy,
\begin{equation}
\frac{ \delta T}{m v_{ts}^2} = \bar{T} = {1 \over 3} (\bar{T}_\parallel + 2 \bar{T}_\perp) = {1 \over 3} \int d^3{\bf v} \, \left[ (v_\parallel^2 - 1) + 2 (\mu B - 1) \right] J_0 h,
\label{coll:energy}
\end{equation}
where again, the Bessel function ultimately arises from the locality
of collisions. Putting it all together, the collision operator is
\begin{equation}
C(h) 
= \nu \left\{ \left[ {\partial \over
     \partial v_\parallel} \left({\partial \over \partial v_\parallel}
   + v_\parallel \right) 
 + 2 {\partial \over \partial \mu} \left( {\mu \over B}
     {\partial \over \partial \mu} + \mu \right) - k_\perp^2 \rho^2
     \right] h  \right. \notag
\end{equation}
\begin{equation}
  \left. \qquad + \left[  \bar{T} \left[
      (v_\parallel^2 - 1) + 2 (\mu B - 1) \right] J_0 +
  {\bf \bar{u}} \cdot {\bf v} \right] F_M \right\}.
\label{eq:fullC}
\end{equation}
The dimensionless collision frequency $\nu$ is defined 
by Eq.~(\ref{coll_freq}). 

This collision operator is a good physical model of like-particle
collisions, which are important to gyrokinetic dynamics.  It captures
the physics of the collision operator presented in \cite{Abel08},
except that our collision frequency does not have velocity
dependence.

\subsubsection{Laguerre-Hermite projection}
This collision operator has an additional attractive feature:
the Laguerre and Hermite polynomials are its eigenfunctions. 
It can therefore be evaluated efficiently. Applying the parallel
velocity components of the cylindrical velocity-space Laplacian
in Eq.~(\ref{eq:fullC}) to the Hermite basis functions gives
\begin{equation}
{\partial \over \partial v_\parallel} \left( {\partial \over \partial v_\parallel}
   + v_\parallel \right) \phi^m
= 
{\partial \over \partial v_\parallel} \left( \phi'^m +
  \sqrt{m+1} \phi^{m+1} + \sqrt{m} \phi^{m-1} \right)
= 
 - m \, \phi^m.
\end{equation}
This means that the Hermite basis functions are eigenfunctions of the parallel
velocity components of the cylindrical velocity-space Laplacian, with
eigenvalue $-m$. Similarly, for the perpendicular components we have
\begin{equation}
2 {\partial \over \partial \mu} \left( {\mu \over B}
     {\partial \over \partial \mu} + \mu \right) \psi^\ell
= 
{2 \ell \over B} {\partial \over \partial \mu} 
(\psi^\ell + \psi^{\ell-1} ) = -2\ell \, \psi^\ell,
\end{equation}
so the Laguerre basis functions are eigenfunctions of the perpendicular components of the Laplacian. The result is that the projection of the velocity-space Laplacian operating on $h$ is sparse in our basis.
Finally, the conservation terms only project onto the $m=0,\ 1,\ \text{and }2$ Hermite moments;
we leave the details to Appendix \ref{app:conservation}.

 Thus the Laguerre-Hermite projection
of our collision operator is
\begin{align}
C_{\ell,0} = &  - \nu \, (b+2\ell + 0) \, H_{\ell,0} \notag\\&+ \nu \left(\sqrt{b} \, \left(\mathcal{J}_\ell + \mathcal{J}_{\ell-1}\right) \bar{u}_\perp +  
\frac{2}{3} \left[  \ell \mathcal{J}_{\ell-1} + 2 \ell \mathcal{J}_\ell + (\ell+1)\mathcal{J}_{\ell+1}\right]
\left( \bar{T}_\parallel + 2\bar{T}_\perp \right)\right),  \notag \\
C_{\ell,1} = & - \nu \, (b + 2 \ell + 1) H_{\ell,1} + \nu \, \gm{\bar{u}_\parallel}{\ell}, \notag \\
C_{\ell,2} = &- \nu \, (b + 2\ell + 2) \, H_{\ell,2} + \nu \, \frac{\sqrt{2}}{3} \, \mathcal{J}_\ell \left( \bar{T}_\parallel + 2\bar{T}_\perp \right), \notag \\
 C_{\ell,m}=&- \nu (b + 2 \ell +m) H_{\ell,m}, \qquad \qquad (m>2) \label{colls}
\end{align}
with the projections of Eqs. (\ref{coll:mom}-\ref{coll:energy}) given by
\begin{gather}
\bar{u}_\parallel = \int d^3{\bf v} J_0 v_\parallel h = \sum_{\ell=0}^\infty \gm{H_{\ell,1}}{\ell}, \label{eq:ubarHL}\\
\bar{u}_\perp = \int d^3{\bf v} J_1 v_\perp h = \sqrt{b_s} \sum_{\ell=0}^\infty \left(\mathcal{J}_\ell+\mathcal{J}_{\ell-1}\right)H_{\ell,0}, \\
\bar{T}_\parallel = \int d^3{\bf v} J_0 \left(v_\parallel^2-1\right) h = \sqrt{2} \sum_{\ell=0}^\infty \gm{H_{\ell,2}}{\ell}, \\
\bar{T}_\perp = \int d^3{\bf v} J_0 \left(\mu B-1\right) h = \sum_{\ell=0}^\infty \left[\ell \mathcal{J}_{\ell-1}+2\ell\mathcal{J}_{\ell} +(\ell+1)\mathcal{J}_{\ell+1} \right]H_{\ell,0}.
\label{eq:tbarHL}
\end{gather}

\subsection{Summary and closure considerations} \label{sec:closures}

We now have a full set of Laguerre-Hermite spectral equations, given
by Eqs.~(\ref{glmevolve}-\ref{hlmqneut}) with collision terms given by Eq.~(\ref{colls}). 
Solving this gyrokinetic
system with many $(\ell, m)$ moments is rigorously equivalent to
solving the gyrokinetic system of Eqs.~(\ref{gk}-\ref{qneut}) with high
resolution in $(v_\parallel, \mu)$ coordinates. There are natural
advantages to using each representation. We expect that pseudo-spectral
algorithms, which have access to both representations, will have
significant advantages over purely spectral and purely $v$-space
algorithms, and also over Lagrangian (such as particle-in-cell) and
semi-Lagrangian formulations, especially with realistic values of
collisionality. This is because the Laguerre-Hermite formulation expresses
critical conservation laws with the first few moments, while the
higher moments are damped progressively more strongly, since $C \sim -
\nu \, (b + 2 \ell + m)$. Our collision operator reflects this
prioritization, as it expresses the key conservation laws at long
wavelength, but sharply attenuates higher moments at short
wavelengths, as a result of classical diffusion, pitch-angle
scattering, energy diffusion, and slowing down. This short wavelength
attenuation is qualitatively correct.

Of course, in practice one cannot solve an infinite set of
Laguerre-Hermite evolution equations. In the limit of high
Laguerre-Hermite resolution, the system can be simply truncated at
some large $\LL$ and $\MM$, and closed by setting
$G_{\ell,m}= H_{\ell,m}=0$ for all perturbations with either
$\ell>\LL$ or $m>\MM$ that appear in the equations for the resolved
moments, where we also take $\mathcal{J}_\ell=0$ for $\ell>\LL$ for
consistency.  We will term this closure approach `closure by
truncation.'  In this case, the unresolved moments correspond to
comparable limitations on any discrete $v$-space representation of
$g(v)$, where fine scales in $g(v)$ above the grid resolution cannot
be resolved. The collision operator regulates these fine velocity
scales by smoothing the distribution function. Our collision operator
fulfills this purpose by acting increasingly strongly on higher
Laguerre and Hermite moments, limiting their amplitude.  Thus for a
given collisionality, there is a physical cutoff at some $\LL$ and
$\MM$ beyond which fine scales in velocity space are completely wiped
out by collisions, which justifies truncation of the moment
series. This is the simplest high-resolution closure, but not the only
option. One can also obtain an asymptotically correct collisional
closure by assuming the collision term becomes dominant in the
unresolved moment equations \cite[]{loureiro2016viriato}.

In the limit of low Laguerre-Hermite resolution, the closure situation is more complicated. 
Unresolved moments are not expected to be negligible at collisionalities of interest, 
so closure by truncation at low resolution will generally give poor results. 
One possible approach is to follow the gyrofluid closure approach pioneered by 
\cite{HP}, which was used and extended in the gyrofluid models of 
 \cite{D1}, \cite{BeerGF}, and \cite{Snyder01}. Further, \cite{Smith97} extended
the Hammett-Perkins approach to an arbitrary number of moments in the slab limit. Thus in the slab limit,
one could use Smith's scheme to generate closures for the Hermite moments. No linear closure is needed for the
Laguerre moments in the slab limit, so in this limit the linear closure problem is solved.\footnote{
Nonlinear phase mixing \cite[]{D1,schekochihin2009astrophysical} does couple Laguerre moments, even in the slab limit, and thus will require closure in some cases. A nonlinear phase mixing closure was included in Dorland's original slab gyrofluid model.}

Extending Smith's generalized closure approach to toroidal geometry is
particularly challenging due to the presence of branch cuts in the
kinetic dispersion relation. In particular, closures for the Hermite
moments are complicated by the fact that phase mixing arises from both
Landau damping and the curvature drift resonance.  Thus we leave the
task of developing generalized closures in toroidal geometry for later
work.  We do however discuss closures that correspond to Beer's
toroidal gyrofluid equations in Appendix \ref{appendix:beer_closures}. This
allows us to reproduce Beer's results using our generalized
Laguerre-Hermite model.

\section{Free energy} \label{sec:free_energy}
The importance of conserved quantities in turbulence is well
appreciated. In the electrostatic gyrokinetic formalism, the (normalized) free
energy defined by 
\begin{align}
W &= \sum_s \int d^3 {\bf r} \int d^3{\bf v} \frac{n_s \tau_{s} \gyavgr{\df_s^2}}{2 F_{Ms}} = \sum_s \int d^3{\bf r} \int d^3{\bf v} \frac{n_s \tau_{s} \gyavgr{h_s^2}}{2 F_{Ms}} - \sum_s \int d^3{\bf r} \frac{n_s Z_s^2}{2 \tau_{s}} \Phi^2 \notag \\
& = \sum_s \int d^3{\bf R} \int d^3{\bf v} \frac{n_s \tau_{s}\ {h_s^2}}{2 F_{Ms}}\ - \sum_s \int d^3{\bf r} \frac{n_s Z_s^2}{2 \tau_{s}} \Phi^2 \label{Wfree}
\end{align}
is conserved in the absence of drive and damping
\cite[]{krommes1993general, Howes06}. 
The free energy evolves in time according to 
\begin{align}
\pderiv{W}{t} &= 
  \sum_s \int d^3 {\bf r} \int d^3{\bf v}  \,
  \frac{n_s \tau_s}{F_{Ms}} \, \gyavgr{h_s \pderiv{g_s}{t}} =  \sum_s \int d^3 {\bf R} \int d^3{\bf v}  \,
  \frac{n_s \tau_s}{F_{Ms}} \, {h_s \pderiv{g_s}{t}}
  \label{Wdot}
\end{align}
Thus, given Eq.~(\ref{gk}) for each species, one calculates 
$\partial W/\partial t$ by multiplying each species' gyrokinetic equation by 
$n_s \tau_s h_s/F_{0s}$, 
integrating the result over the entire phase space, and summing over species.

We will now explore the conservation of $W$ in the Laguerre-Hermite
basis. Expanding and projecting $h_s$ in Eq.~(\ref{Wfree}), we have
\begin{align}
W &= \sum_s n_s \tau_s \int d^3 {\bf R} \sum_{\ell=0}^{\LL} \sum_{m=0}^{\MM} 
  \frac{1}{2} H_{\ell,m}^2
   - \sum_s \frac{n_s Z_s^2}{2 \tau_s} \int d^3{\bf r}\, \Phi^2 \label{Wfreedot}
\end{align} 
where for now we have truncated the Laguerre-Hermite expansion at some $\LL$ and $\MM$. 
We will primarily address free energy conservation under the assumption of closure by truncation,
as described in Section \ref{sec:closures}.\footnotemark\ 
We can also recover the fully kinetic limit by taking $\LL,\MM\rightarrow\infty$.

\footnotetext{The situation is again more complicated when using more sophisticated closures
in the style of Hammett \& Perkins. In this light, we discuss some subtleties of free energy conservation in 
the Beer gyrofluid model in Appendix \ref{appendix:beer_free_energy}.}

The free energy for each species $W_s$ evolves according to
\begin{align}
\pderiv{W_s}{t} &=n_s \tau_s \int d^3 {\bf R} \left[ \sum_{\ell=0}^{\LL}
  \sum_{m=0}^{\MM} H_{\ell,m} \pderiv{G_{\ell,m}}{t} \right].
\label{WHG}
\end{align}
Most of the terms in this large sum cancel, leaving only contributions
that involve perturbations with either $\ell>\LL$ or $m>\MM$. These
remainder contributions are precisely where closures are required.

Since there are several different types of terms in the $\Gjk$
evolution equations, it is convenient to consider the free energy
evolution of each separately. The parallel convective derivative terms
(including the $\nabla_\parallel \ln B$ magnetic trapping terms) give
\begin{align}
\frac{1}{n_s \tau_s v_{ts}} \pderiv{W_\parallel}{t} 
= &- \int d^3 {\bf R} \left[\sum_{\ell=0}^\LL \sqrt{\MM+1} H_{\ell,\MM} \nabla_\parallel H_{\ell,\MM+1}\right] \notag
\\\quad+ &\left[\sum_{\ell=0}^\LL
  \left[-(\ell+1)\sqrt{\MM+1}H_{\ell,\MM}H_{\ell,\MM+1} - \ell \sqrt{\MM+1} H_{\ell,\MM}H_{\ell-1,\MM+1}\right]
  \nabla_\parallel\ln B\right]\notag\\+ &\left[ \sum_{m=0}^\MM \left[\sqrt{m} (\LL+1) H_{\LL,m}H_{\LL+1,m-1} \right]
  \nabla_\parallel\ln B\right] 
, \label{Wpar}
\end{align}
where we have used the identity 
\begin{align}
\nabla_\parallel (F G) = B \nabla_\parallel \left(\frac{F G}{B}\right) + FG\nabla_\parallel \ln B
\end{align}
repeatedly, and also used the fact that $B \nabla_\parallel(FG/B)$ has
the form of a total divergence, which vanishes upon integration over
all space (keeping in mind that $B$ is the Jacobian of our coordinates).

Each of these terms describes the coupling of one resolved
Laguerre-Hermite moment $(\ell \leq \LL, m \leq \MM)$ to one unresolved
moment. In the first sum on the right hand side, there are $\LL+1$
unresolved moments, each with $m=\MM+1$; each of these moments has
contributions from $\int d^3{\bf v}\, v_\parallel^{\MM+1} v_\perp^{2\ell} g$; all
relate to the {\it largest} scale in $g(v_\parallel)$ that {\it
  cannot} be resolved. There are comparable limitations on any
discrete $v$-space representation of $g(v_\parallel)$. 
Similarly, in the second and third sums on the right hand side, there are $(\MM+1)$
and $(\LL+1)$ additional unresolved moments that must be closed, respectively. When closing by
truncation, we take $H_{\ell,\MM+1}=H_{\LL+1,m-1}=H_{\ell,\MM+1}=H_{\ell-1,\MM+1}=0$, so
all terms in these three sums vanish. 
Thus closure by truncation gives $\partial W_\parallel/ \partial {t}=0$.

Note that in the kinetic limit, $\partial{W_\parallel}/\partial{t}\rightarrow0$
as long as $G_{\LL,\MM}\rightarrow0$ ``fast enough'' as
$\LL,\MM\rightarrow\infty$, which is the expected result.
This will be the case for all of the parts of the free energy except the drive and damping parts.

The toroidal drift terms give
\begin{align}
\frac{Z_s}{n_s \tau_s^2} \pderiv{W_{d}}{t} 
&= - \int d^3{\bf R} \Bigg[ \sum_{\ell=0}^\LL \sqrt{(\MM+1)(\MM+2)} H_{\ell,\MM}i\omega_d H_{\ell,\MM+2}
   \notag \\&+ \sqrt{\MM (\MM+1)} H_{\ell,\MM-1} i\omega_d H_{\ell,\MM+1}\Bigg] + \ \sum_{m=0}^{\MM} (\LL+1) H_{\LL,m} i\omega_d H_{\LL+1,m}, 
\end{align}
where here $i\omega_d(F G)$ has the form of a total divergence and
vanishes upon integration over all space. There are $2(\LL+1)$
unresolved moments in the first sum, and $(\MM+1)$ unresolved moments in the
second sum. As before, closure by truncation ensures that all of these terms
vanish, giving $\partial W_d / \partial t = 0$. 

The nonlinear terms give
\begin{align}
\frac{1}{n_s \tau_s} \pderiv{W_{NL}}{t} 
&= -\int d^3 {\bf R} \sum_{\ell=0}^\LL \sum_{m=0}^\MM H_{\ell,m}\sum_{k=\LL+1}^\infty \sum_{n=|k-\ell|}^{k+\ell} 
   \alpha_{k\ell n} \gm{}{n}\vE \cdot \nabla H_{k,m}.
\end{align}
This remainder term describes nonlinear coupling of resolved moments $H_{\ell<\LL+1, m}$ to
unresolved moments $H_{k>\LL, m}$ via FLR corrections to the ${\bf
  E}\times{\bf B}$ drift. Once again, closure by truncation ensures that all of these terms
vanish, giving $\partial W_{NL} / \partial t = 0$.

For appropriate choices of density
and temperature gradients, the drive terms can cause the
free energy to increase:
\begin{align}
\frac{1}{n_s \tau_s} \pderiv{W_{*}}{t} 
&= \int d^3 {\bf R} \sum_{\ell=0}^\LL \Bigg[ H_{\ell,0} i \omega_* 
   \left( \frac{1}{L_n} \gm{\Phi}{\ell} + \frac{1}{L_T} \big(\ell \gm{\Phi}{\ell-1} + 2\ell \gm{\Phi}{\ell} + (\ell+1)\gm{\Phi}{\ell+1}\big)\right) \notag \\
   &\qquad\qquad\qquad+ \frac{1}{\sqrt{2}}\frac{1}{L_T} H_{\ell,2}
     i\omega_* \gm{\Phi}{\ell}\Bigg] \notag \\
   & = \int d^3 {\bf r} \left( \frac{\bar{n}}{L_n} + \frac{\bar{T}}{L_T} \right) i
     \omega_* \Phi \quad \neq \quad 0.
\end{align}
These terms require no closure, though we will take
$\mathcal{J}_{\LL+1}=0$ when closing other terms by truncation for consistency.

Because the Laguerre-Hermite moments are eigenfunctions of the
velocity-space derivatives of our collision operator, each moment
contributes $-\nu (b + 2 \ell + m) H_{\ell,m}^2$ to $dW/dt$. The
field-particle (conservation) terms are more complicated, but as shown
in Appendix \ref{app:collisions}, never result in a net increase $W$ after summing over all
$(\ell, m)$. Thus, the collision operator can only decrease the total
free energy or leave it unchanged:
\begin{align}
  \frac{1}{n_s\tau_s} \pderiv{W_C}{t} & = \int d^3 {\bf R} \int d^3 {\bf v} \, \frac{h}{F_M}\, C(h) \notag \\
  & = -\nu \sum_{\ell=0}^\LL \sum_{m=0}^\MM \left[ (b + 2 \ell + m) |H_{\ell, m}|^2 \right]
  + \nu (|\bar{u}_\parallel|^2 + |\bar{u}_\perp|^2 + 3 |\bar{T}|^2) \notag \\
  & \leq 0.
\label{Wcdot}
\end{align}

Upon denoting the total drive by $\mathcal{D} \equiv \partial W_*/\partial t$
and the total effect of collisions by $\mathcal{C} \equiv -\partial W_C/\partial t$, 
we have
\begin{align}
{dW \over dt} &= \sum_s n_s \tau_s \left(v_{ts} \pderiv{ W_\parallel}{t} + \frac{\tau_s}{Z_s}\pderiv{W_d}{t} +
\pderiv{W_{NL}}{t}  + \mathcal{D} - \mathcal{C} \right),  \\ \lim_{\LL,\MM\to\infty}
{dW \over dt} &=\mathcal{D} - \mathcal{C},
\end{align}
where in the second line we have indicated the desired kinetic result,
that the total free energy of each species $W_s$ for the truncated
system is conserved in the absence of driving and collisional damping. 

\section{Linear Results}
We now show some preliminary linear calculations, as a proof of
concept of our Laguerre-Hermite formulation of gyrokinetics. These
calculations have been performed with \texttt{GX}, a new gyrokinetic
code that uses our Laguerre-Hermite spectral velocity
discretization. Numerical details and further results from \texttt{GX}
will be reported separately.

\begin{figure}
\centering
\begin{minipage}{.45\linewidth}
\hspace{-.5cm}
\includegraphics[width=1.2\linewidth]{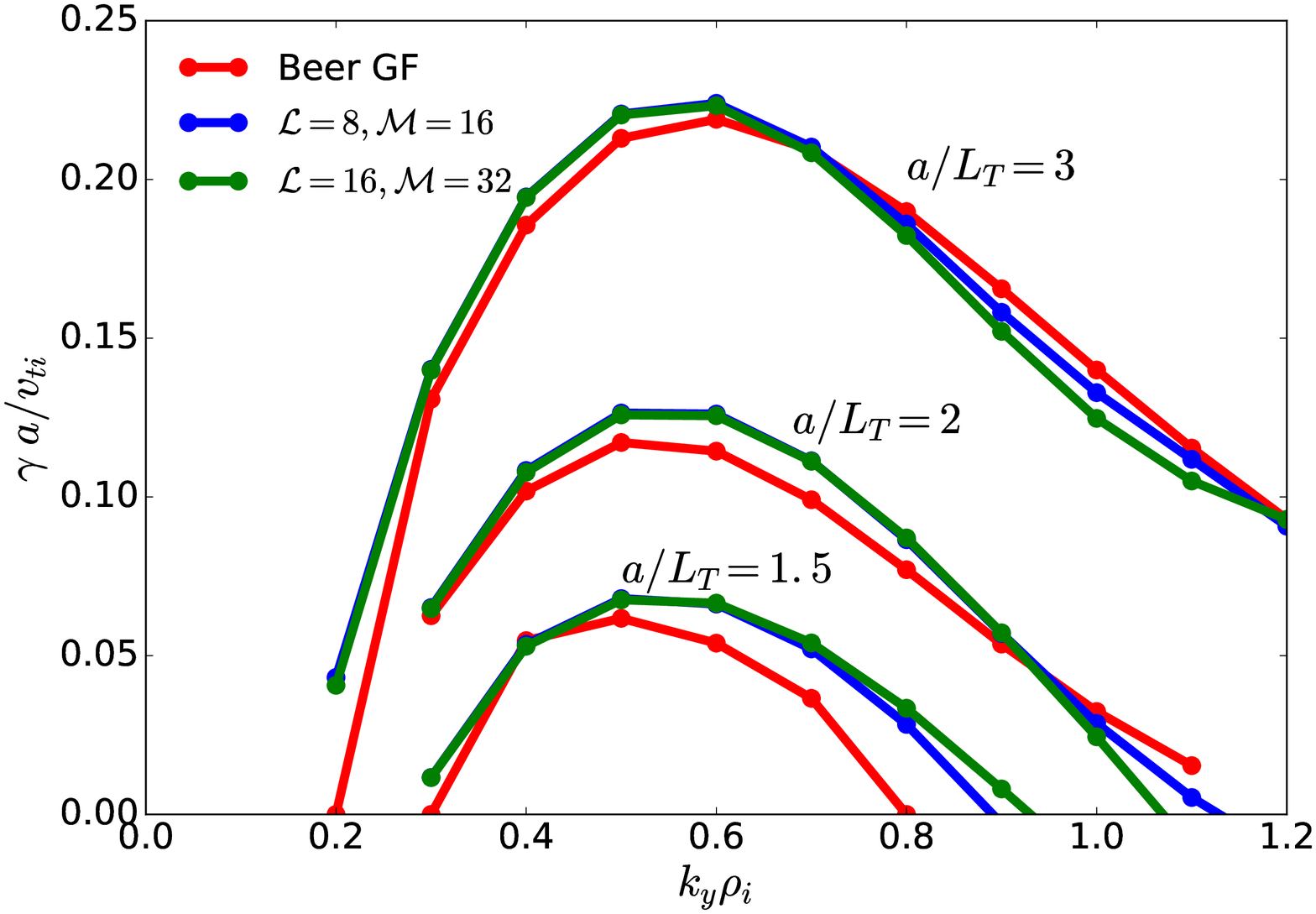}
\caption{Growth rates of an ITG instability in the local limit. The results from the Laguerre-Hermite formulation are shown at two choices of velocity resolution (blue and green), as well results from the Beer 4+2 gyrofluid model (red). The Laguerre-Hermite results show good convergence in velocity resolution, especially at small $k_y \rho_i$.}
\label{fig:local}
\end{minipage}\qquad\qquad
\begin{minipage}{.45\linewidth}
\hspace{-1cm}
\resizebox{!}{2.23in}{\input{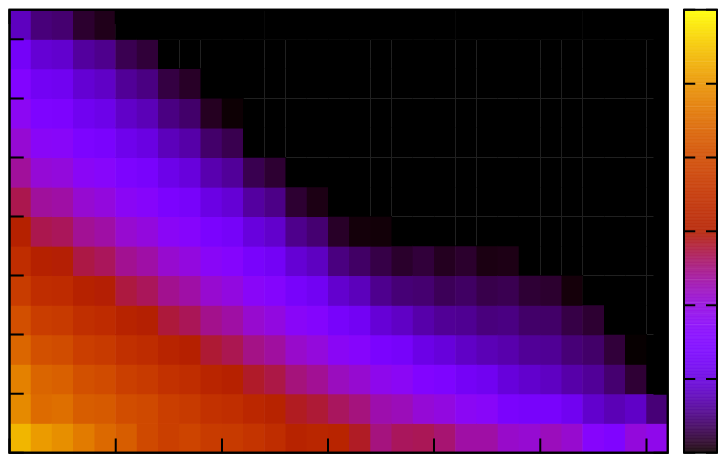}}
\vspace{-.9cm}
\caption{The normalized integrated velocity space spectrum, defined by Eq.~(\ref{spectrum}), resulting from the higher resolution $a/L_T=2$ case in Figure \ref{fig:local}. The amplitudes decrease by several orders of magnitude moving from small to large $\ell$ and $m$, indicating that the calculation is well resolved in velocity space.}
\label{fig:local_hlspectrum}
\end{minipage}
\end{figure}

\subsection{Local Linear ITG}
We first examine an ITG instability in the local limit, where
$k_\parallel$, $B$, and $\omega_d$ are treated as constants. We select
a case used to validate the Beer 4+2 gyrofluid model \cite[]{BeerGF},
which first appeared in \cite{Dong92}. The relevant parameters (in our
units) are $q=2$, $R_0/a=5$, $k_\parallel a = a / q R_0$, $a/L_n=1$,
and $a/L_T=1.5,\ 2,\ \text{and }3$. We also set $\nu=0.01$.  Figure
\ref{fig:local} shows linear growth rates over a range of $k_y \rho_i$
for two choices of Laguerre-Hermite velocity resolution. For
comparison, we also show the results from the Beer gyrofluid model. We
see that the growth rates converge as the Laguerre-Hermite resolution
is increased. Note that these calculations use closure by truncation.

We calculate a normalized integrated velocity space spectrum, 
\begin{equation}
P(\ell,m) = \frac{\int d^3 {\bf R}\ |G_{\ell,m}|^2 }{  \int d^3 {\bf R}\ |G_{0,0}|^2}, \label{spectrum}
\end{equation}
to examine the structure of the distribution function in the
Laguerre-Hermite basis. $P$ can also be interpreted as a measure of
the free energy in each moment. Figure \ref{fig:local_hlspectrum}
shows this spectrum for the higher resolution case in the above
$a/L_T=2$ calculation. The amplitude $P$ is highest at small $\ell$
and $m$, which is expected since free energy is injected by the
gradients at these large scales. 
The upper
right quadrant of the plot, where both $\ell$ and $m$ are large, is
more than six orders of magnitude smaller than the large amplitudes at
small $\ell$ and $m$. For best contrast we have truncated the color
scale to seven orders of magnitude, but note that the amplitude in the
top right corner, corresponding to the finest resolved scale in the
velocity space, is $P(15,31)\sim10^{-10}$. This is a good indication
that this calculation is well resolved in velocity space. Further, we
see that $P(7,15)\sim10^{-5}$. This is the finest resolved scale in
the lower resolution case, which has a spectrum (not shown) that looks
qualitatively similar to the lower left quadrant of Figure
\ref{fig:local_hlspectrum}, and which also has
$P(7,15)\sim10^{-5}$. The fact that the lower resolution case agrees
with the higher resolution case in both growth rates and spectra
suggests that the low resolution case has sufficient resolution. 

\begin{figure}
\begin{minipage}{.45\linewidth}
\hspace{-.5cm}
\includegraphics[width=1.2\linewidth]{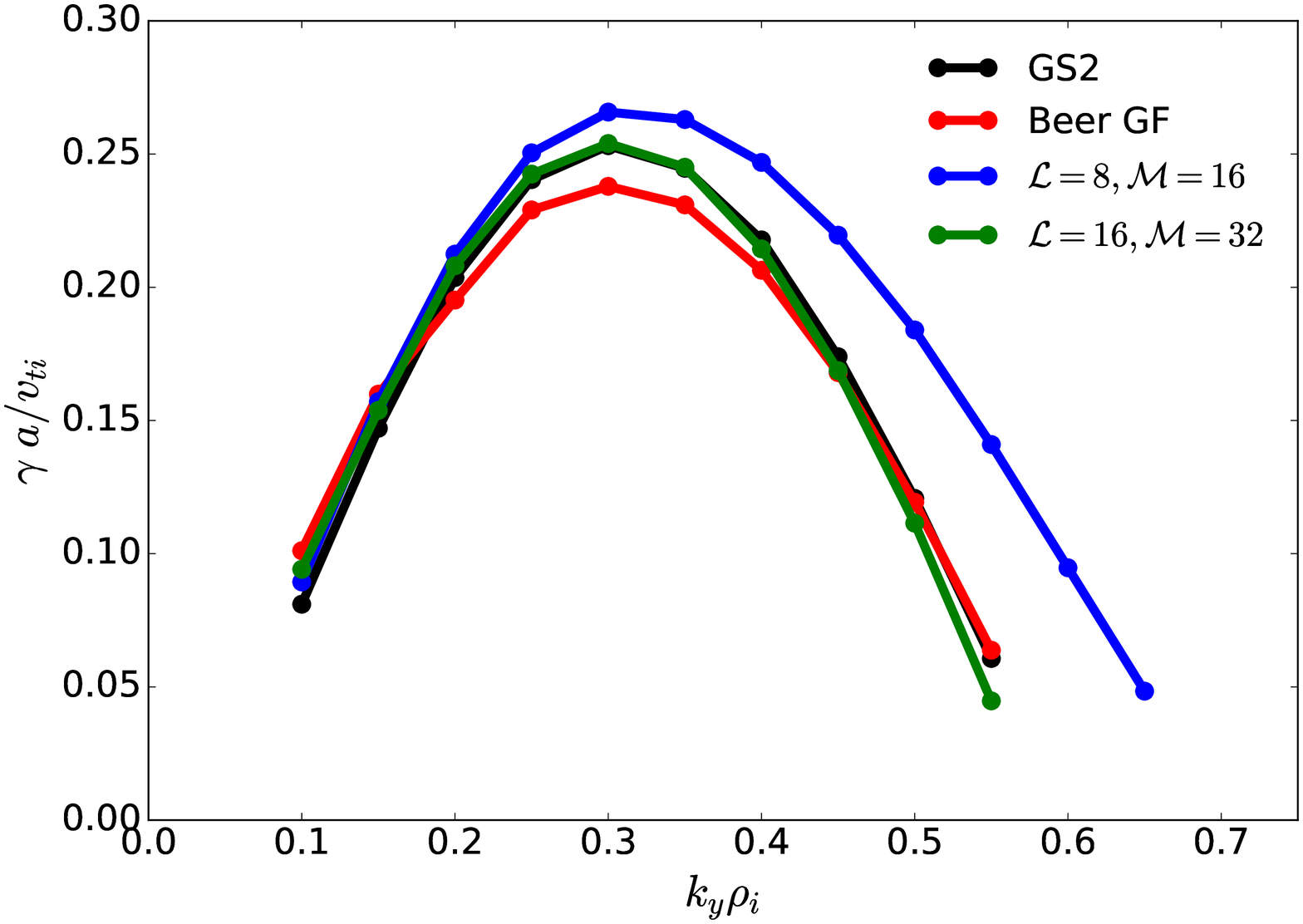}
\caption{Linear growth rates for the Cyclone base case \cite[]{Cyclone}. The results from the Laguerre-Hermite formulation are shown at two choices of velocity resolution (blue and green), as well as results from the gyrokinetic code \texttt{GS2} (black) and the Beer 4+2 gyrofluid model (red).}
\label{fig:cyclone}
\end{minipage}\qquad\qquad
\begin{minipage}{.45\linewidth}
\hspace{-1cm}
\resizebox{!}{2.23in}{\input{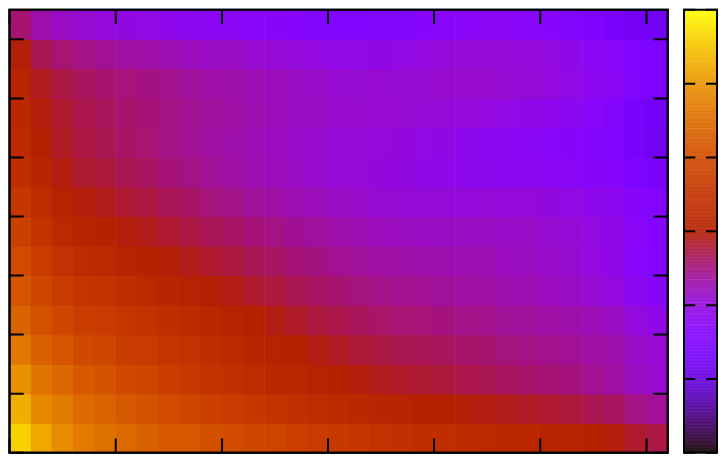}}
\caption{The normalized integrated velocity space spectrum resulting from the higher resolution case in Figure \ref{fig:cyclone}. While this spectrum is still well resolved, the amplitudes decrease more gradually than in the local limit spectrum from Figure \ref{fig:local_hlspectrum}.}
\label{fig:cyclone_hlspectrum}
\vspace{1cm}
\end{minipage}
\end{figure}

\subsection{Cyclone Linear ITG}
We also examine an ITG instability in the nonlocal limit, for which we
use the Cyclone base case parameters, a widely benchmarked test case
\cite[]{Cyclone}. Figure \ref{fig:cyclone} shows the linear growth
rates over a range of $k_y \rho_i$ for the same two choices of
Laguerre-Hermite velocity resolution as above. For comparison, we also
show the results from the Beer 4+2 gyrofluid model, along with results
from the gyrokinetic code \texttt{GS2}. \texttt{GS2} solves the
gyrokinetic equation using a polar velocity grid in energy and pitch
angle coordinates, $\varepsilon$ and $\lambda = \mu/\varepsilon$. In
the limit of large velocity resolution, our Laguerre-Hermite approach
should agree with the grid-based approach of \texttt{GS2}. From the
figure we see that the two approaches do indeed agree in the higher
resolution case. We also see that the convergence in resolution is
slower here in the nonlocal limit than in the local limit above, as
the lower resolution case gives growth rates too large for higher
$k_y \rho_i$. This suggests that more velocity resolution is needed in
the nonlocal limit than in the local limit.

This is confirmed by the velocity spectrum for this case, shown in
Figure \ref{fig:cyclone_hlspectrum}. Whereas in the local limit a
significant portion of the velocity space had amplitudes more than six
orders of magnitude smaller than the largest amplitudes, in the
nonlocal limit the amplitudes decrease much more gradually. This
indicates that the distribution function has more structure in
velocity space in this case. This also shows why the lower resolution
case failed to produce accurate growth rates: the amplitudes in the
lower left quadrant of Figure \ref{fig:cyclone_hlspectrum} do not
decrease by as much as in the local case, with $P(7,15)\sim10^{-4}$.

\subsection{Rosenbluth-Hinton Zonal Flow Residual}
As a final test, we examine zonal flow dynamics in our
Laguerre-Hermite formulation. Zonal flows are nonlinearly-driven and
nonlinearly-damped \cite[]{Rogers00} toroidally and poloidally
symmetric sheared ${\bf E} \times {\bf B}$ flows that have been shown
to play a key role in determining the turbulence saturation level
\cite[]{H_sher93, Waltz94}. \cite{RH} first showed that
zonal flows are not linearly damped by collisionless processes,
showing that in a simplified equilibrium model the residual flow is
given by
\begin{equation}
\label{RH}
\frac{\Phi(t=\infty)}{\Phi(t=0)} = \frac{1}{1+1.6q^2/\sqrt{\epsilon}},
\end{equation}
where $q$ is the safety factor and $\epsilon=r/R_0$ is the inverse
aspect ratio. The original Beer gyrofluid model did not accurately
capture the residual, causing discrepancies with gyrokinetic models in
nonlinear simulations \cite[]{Cyclone}. The gyrofluid closure was
modified to be able to capture some of the zonal flow residual with
limited success \cite[]{Varenna98}.

\begin{figure}
\begin{minipage}{.45\linewidth}
\hspace{-.5cm}
\includegraphics[width=1.2\linewidth]{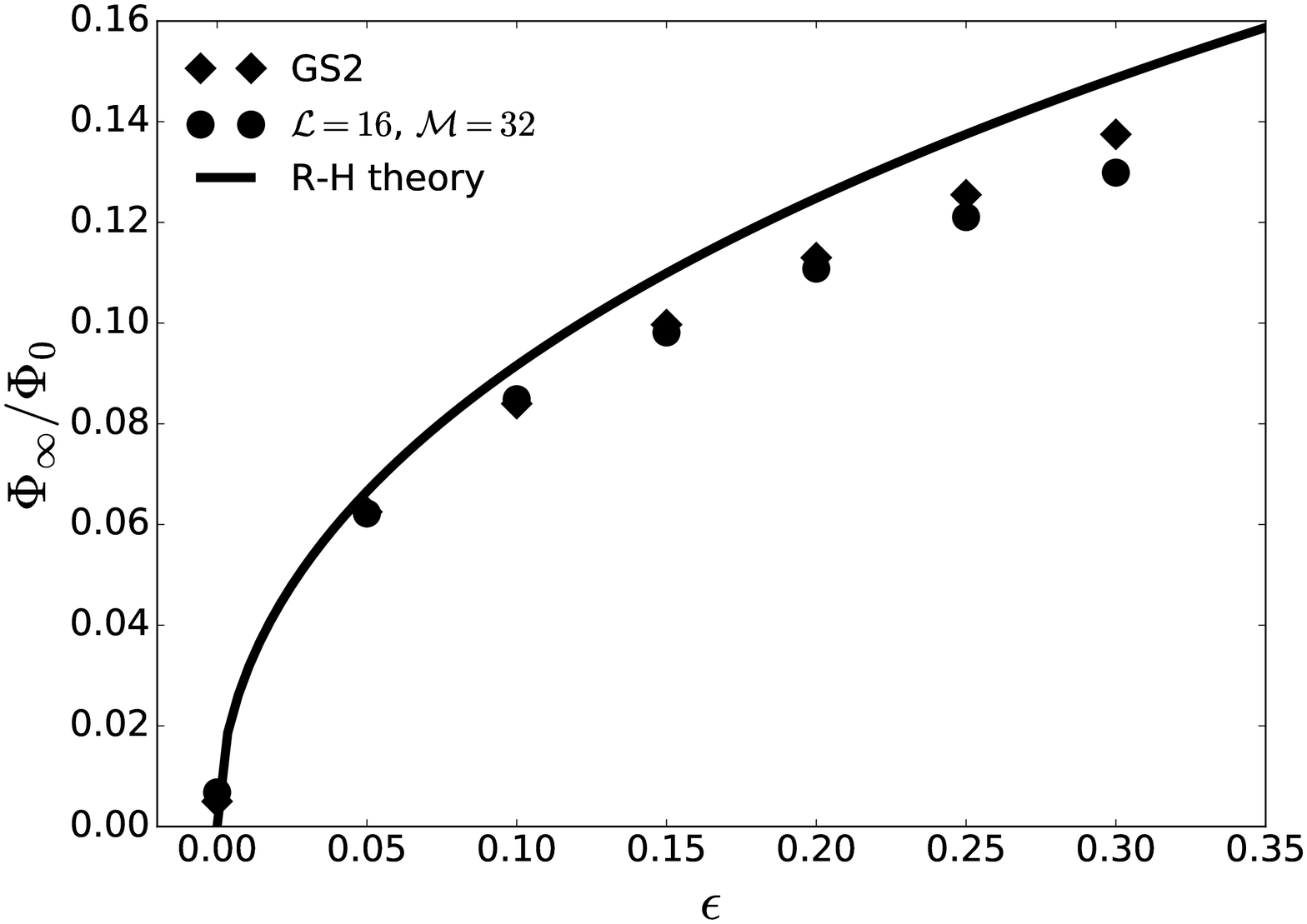}
\caption{Rosenbluth-Hinton residual flow for various values of $\epsilon=r/R_0$.
The residuals calculated by the Laguerre-Hermite model with $\LL=16$ and $\MM=32$
agree well with those calculated by \texttt{GS2}. We also show the expected theoretical
result, Eq. (\ref{RH}).}
\label{fig:RH}
\end{minipage}\qquad\qquad
\begin{minipage}{.45\linewidth}
\hspace{-1cm}
\resizebox{!}{2.23in}{\input{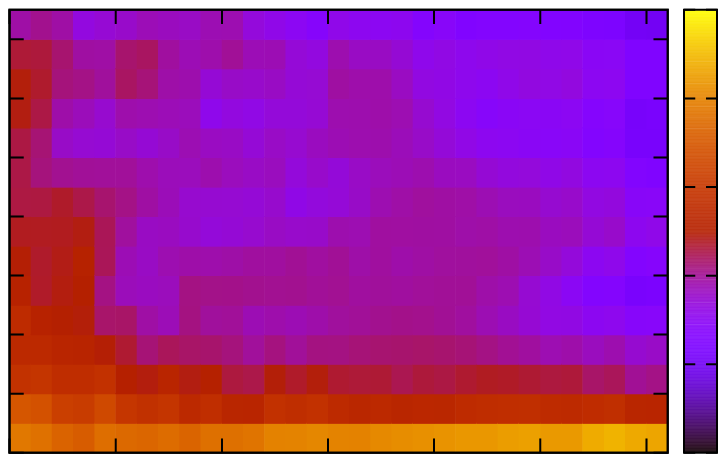}}
\caption{The normalized integrated velocity space spectrum for the $\epsilon=0.2$ case from
Figure \ref{fig:RH}. High amplitudes at large $m$ along the axis are the result of trapped particle
dynamics that produce sharp features in $v_\parallel$.}
\label{fig:RH_hlspectrum}
\vspace{.6cm}
\end{minipage}
\end{figure}

Figure \ref{fig:RH} shows residuals for several values of $\epsilon$, for the $k_x \rho_i = 0.01$ mode with $q=1.4$, as calculated by our Laguerre-Hermite model with $\LL=16$ and $\MM=32$. 
We also show the residual calculated by \texttt{GS2} and the expected 
theoretical value from Eq.~(\ref{RH}). 
Our result for the average residual agrees well with \texttt{GS2}, 
though we note that we observe larger oscillations and less
damping of the potential in the Laguerre-Hermite model. We attribute this to resolution
issues. The velocity spectrum, shown in Figure \ref{fig:RH_hlspectrum} for the $\epsilon=0.2$ case, shows overall larger amplitudes compared to the spectra from the
linear instability calculations. 
We also see that there are high amplitudes at large $m$ along the axis, indicating
that sharp features are being generated in $g(v_\parallel)$. 
This is to be expected, since the zonal flow residual is the result of complicated
kinetic effects involving trapped particles dynamics, and resolving the trapped-passing boundary (which the velocity grid in  \texttt{GS2} is explicitly designed for)
is important. Naturally, our spectral approach is not as well suited
to resolving these sharp features.
In the problems we will target, however, collisions and nonlinearity will 
help to smear away sharp features in the  
distribution function.
 

\section{Summary \& Conclusion}

In this report, we have outlined a new pseudo-spectral velocity
formulation of flux-tube gyrokinetics. This is achieved by projecting
the gyrokinetic equation onto a velocity basis composed of Laguerre
and Hermite polynomials. A key advantage of the resulting model is the
flexibility in choice of velocity-space resolution.  At the lowest
velocity-space resolution, the model corresponds directly to gyrofluid
models. Within the same framework, the model corresponds to
conventional gyrokinetic approaches at high velocity resolution.
Between these limits, one has the freedom to tailor the resolution to
one's needs: one can smoothly increase resolution to improve accuracy,
or one can minimize resolution to produce a performance advantage.
There is also an opportunity for a dynamic fidelity refinement
approach, where the resolution can be adjusted dynamically during a
simulation.

The flexibility to use relatively lower velocity resolution than is
used in standard gyrokinetic approaches is due to the fact that our
Laguerre-Hermite formulation expresses critical conservation laws even
at very low resolution.  We have presented a model collision operator
that reflects this prioritization: it expresses the key conservation
laws at long wavelength, but sharply attenuates higher moments at
short wavelengths as a result of classical diffusion, pitch-angle
scattering, energy diffusion, and slowing down. Our collision operator
is also efficiently expressed in our basis, and it satisfies the H
theorem.

The main disadvantage of our approach is the need for closures due to
the coupling of the spectral moments, which arises as the mathematical
manifestation of the physics of phase mixing.  We recognize that the
value of our approach will only be fully realized when a generalized
closure scheme has been found that can give gyrokinetic fidelity with
sub-kinetic velocity resolution. The relative success of gyrofluid
closures indicates that this is feasible, and we can use these
closures directly in our model at the lowest resolution (see Appendix
\ref{appendix:beer_closures}).  We thus leave as important future work
the generalization of the closures to arbitrary spectral velocity
resolution. This will bolster the flexibility of the model, as one
will be able to capture important kinetic effects with whatever
fidelity is required by the problem at hand. This includes modeling
linear phase mixing from Landau damping and toroidal drifts as well as
nonlinear effects such as nonlinear FLR phase mixing
\cite[]{schekochihin2009astrophysical, tatsuno2009nonlinear, PlunkJFM,
numata2015ion, howes2011gyrokinetic, chen2010interpreting,
kawamori2013experimental, D1}, and phase
un-mixing from the plasma echo \cite[]{schekochihin2015phase,
  kanekar2014phase, parker_dellar_2015}.

We also leave the extension of this model to include electromagnetic
fluctuations for the future. While we could simply include electrons
as a second species and find the perpendicular magnetic field
fluctuations from Ampere's law, such an approach is potentially
expensive. For many applications, the \cite{Snyder01} model
will be the best way forward, for example. Importantly, a model with 
$\delta A_\parallel$ and $\delta B_\parallel$ presents no new
challenges to the Laguerre-Hermite projections. 

Our preliminary results show that our model is capable of reproducing
gyrokinetic results for linear instabilities and zonal flow dynamics
at spectral resolution comparable to conventional gyrokinetic velocity
resolution. Nonlinear simulations of ion temperature gradient
turbulence, which make use of the pseudo-spectral approach described
in Appendix \ref{app:HLeval}, will be reported in a later work. We
have also coupled our model to the TRINITY full-torus gyrokinetic
transport framework \cite[]{Barnes10}. In this setting the flexibility
of our model enables a significant performance advantage over standard
approaches, allowing efficient and tractable whole-device modeling of
tokamaks and stellarators \cite[]{Highcock16}.

Finally, we note that our model is inherently parallelizable and small
memory footprint due to its pseudo-spectral approach. This allows us
to move a well-resolved gyrokinetic simulation of ion temperature
gradient-driven tokamak turbulence from hundreds or thousands of cores
on an massively parallel HPC device to a single graphics processor
(GPU). Thus our new Laguerre-Hermite pseudo-spectral formulation has
produced a high-fidelity gyrokinetic simulation code that runs at a
few teraflops on a desktop computer with an inexpensive GPU.
\\

We would like to thank Ian Abel, Per Helander, Edmund Highcock, Greg
Hammett, Alex Schekochihin, Michael Barnes, and Kate Despain for
fruitful discussions and encouragement. Research support came from the
U.S. Department of Energy: NRM is supported by the DOE CSGF program, provided under grant
DE-FG02-97ER25308; {WD is supported by award numbers
  DE-FC02-08ER54964 and DE-FG02-93ER54197}; ML is supported by award number
DE-FG02-93ER54197.  Computations were performed at the Texas Advanced
Computing Center (TACC) at The University of Texas at Austin, via the Extreme
Science and Engineering Discovery Environment (XSEDE), which is
supported by National Science Foundation grant number ACI-1548562.

\appendix 

\section{Non-dimensionalization} \label{app:norm}

In $(v_\parallel, \mu)$ coordinates, the gyrokinetic equation can be
written \cite[]{Brizard,BeerGF,Snyder01, parra2015equivalence} as: 
\begin{align}
\pderiv{g}{t} & + 
\left[ v_\parallel \bhat + \langle \vE \rangle_{\bf R} + {\bf v}_d \right] 
\cdot \nabla h
+ \langle \vE \rangle_{\bf R} \cdot \nabla F_M
 - \mu (\bhat\cdot\nabla B) \pderiv{h}{v_\parallel} 
= C(h).
\label{eq:gk}
\end{align}
In this equation, the distribution function
$F = F_0 + \df = F_M (1 - q \Phi/T) + h$, and the perturbed,
gyroaveraged distribution function
$g = \langle \df \rangle_{\bf R} = h - q \langle \Phi \rangle_{\bf R}
F_M / T$, where $\langle \dots \rangle_{\bf R}$ denotes a gyroaverage
at fixed guiding center position ${\bf R}$, and $F_M$ is a Maxwellian
distribution function.

Radial gradients of the equilibrium distribution function are assumed to vary on a
scale $a$ (which may be taken to be the minor radius, major radius, or
some other convenient macroscale length). All equilibrium-scale
lengths, such as density gradient scale lengths, magnetic curvature,
and so on, are normalized by $a$. The magnetic field $B$ is normalized
by $B_a$, which may be chosen for convenience. A typical choice would
be the vacuum magnetic field at the center of the last closed flux
surface, at the elevation of the magnetic axis. 

Fluctuating quantities are assumed to vary along the field line slowly
and across the field line rapidly. We thus normalize
variations of fluctuations along the field line by $a$, and across
field lines by $\rho_{\rm ref}$, where $\rho_{\rm ref}$ is the thermal
gyroradius of a convenient reference species, which is in turn
characterized by its charge $q_{\rm ref}$, mass $M_{\rm ref}$,
temperature $T_{0, {\rm ref}}$, density $n_{0, {\rm ref}}$. Throughout
this paper, we define thermal speeds by $v_t \equiv \sqrt{T/M}$, so
that 
$$
\rho_{\rm ref} = v_{t, {\rm ref}} / \Omega_{\rm ref} = 
\left(\sqrt{\frac{T_{0,  {\rm ref}}}{M_{\rm ref}}}\right) \frac{M_{\rm ref} c}{q_{\rm ref} B_a}.
$$
In this expression, $c$ represents the speed of light. 

The gyrokinetic expansion parameter $\varepsilon$ can be taken to be
$\varepsilon = \rho_{\rm ref} / a \equiv \rho_*$. Fluctuations vary slowly in time
compared to the gyrofrequency. Accordingly, the fluctuation time scale is
normalized by $\Omega_{\rm ref} \rho_* = v_{t, {\rm ref}} /
a$. Similarly, fluctuating quantities are small compared to
equilibrium quantities ({\it e.g.,} fluctuating density $\delta n \sim
\varepsilon n_0$. The equations are therefore most naturally expressed
by scaling the fluctuating quantities by $1/\rho_{\rm ref}$. This
includes the electrostatic potential, which is normalized by $T_{0, {\rm ref}} /
q_{\rm ref}$. That is, the physical electrostatic potential $\Phi_{\rm
  phys} = (T_{0, {\rm ref}} / q_{\rm ref}) \tilde{\Phi} \rho_*$, where
$\tilde{\Phi}$ is the normalized fluctuating electrostatic potential. 

For each species, we normalize the velocity space coordinates
$v_\parallel$ and $\mu$ using the thermal velocity for that
species. This thermal velocity is in turn normalized to the thermal
velocity of the reference species. Thus,
$v_\parallel \bhat \cdot \nabla \rightarrow (v_\parallel / v_{t, s}) (v_{t,
  s}/v_{t, {\rm ref}}) a \tilde{\nabla}_\parallel$. In the body of
this report, we simplify the final equations by cancelling common
factors and by dropping the tilde notation. Thus, this operator would
appear in explicit form as $v_{ts} v_\parallel \bhat \cdot \nabla$,
where $v_{ts} \equiv v_{t, s}/v_{t, {\rm ref}}$, and it would be
understood from context that $v_\parallel$ had been normalized
appropriately for species $s$, and that $\bhat \cdot \nabla$ had been
normalized to the length $a$. When the context is clear, we may also
neglect to write the leading coefficient $v_{ts}$. 

The non-dimensional collision frequency $\nu_s$ is given by 
\begin{equation}
\nu_s = \frac{n_s Z_s^4 v_{ts}}{\tau_s^2} \nu_{\rm ref}, \label{coll_freq}
\end{equation}
where $\nu_{\rm ref}$ is the collision frequency of the reference species
normalized by $a/v_{t, {\rm ref}}$:
\begin{equation}
\nu_{\rm ref} = \frac{\sqrt{2} \pi \ a\ n_{0,{\rm ref}} \ e^4 \ln \Lambda}{T_{\rm ref}^2}.
\end{equation}

The normalized time coordinate is $\tilde{t} = t_{\rm phys} v_{t, {\rm
    ref}} / a$. For species $s$, we define the temperature $\tau_s \equiv T_{0,
  s}/T_{0, {\rm ref}}$, the density $n_s \equiv n_{0, s}/n_{0, {\rm
    ref}}$, the charge $Z_s = q_s/q_{\rm ref}$, the normalized mass
$m_s = M_s/M_{\rm ref}$, and the normalized thermal gyroradius $\rho_s
= (v_{t, s}/\Omega_s) /\rho_*$.  

Fluctuating densities and temperatures are normalized by their
equilibrium values; fluctuations of odd $v_\parallel$ moments are
normalized by additional appropriate powers of $v_{ts}$. The
normalized gyrokinetic equation [Eq.~(\ref{gk})] results. 
Elsewhere, we restore normalizations and species
subscripts only when needed for clarity.

\section{Linear FLR accuracy} \label{appendix:flr}

\begin{figure}
\captionsetup[subfigure]{justification=centering}
\begin{subfigure}{.45\linewidth}
\begin{center}
\includegraphics[scale=.7]{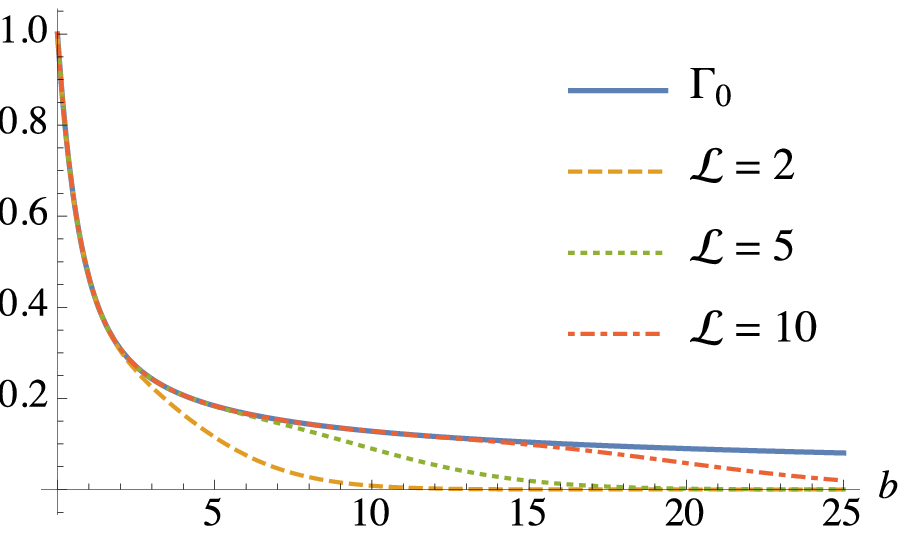}
\label{fig:gamma0}
\end{center}
\end{subfigure}
\qquad   
\begin{subfigure}{.45\linewidth}
\begin{center}
\includegraphics[scale=.7]{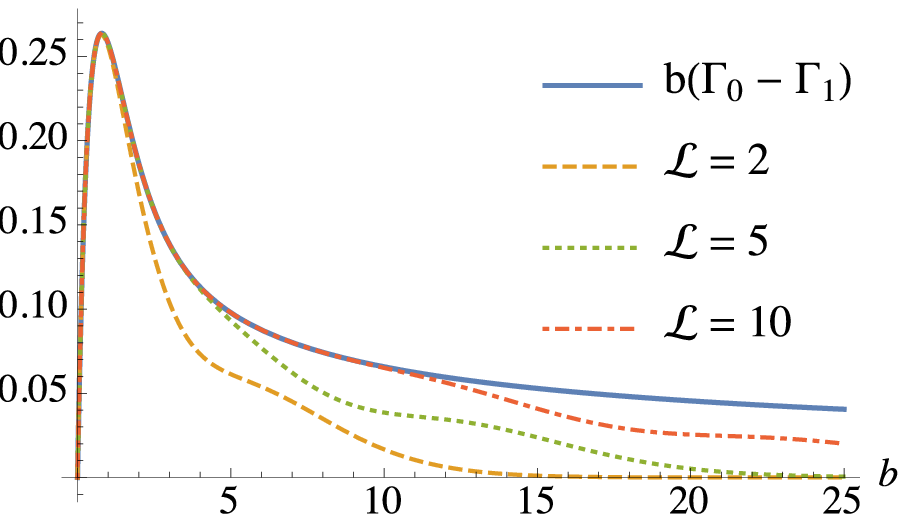}
\label{fig:bgamma}
\end{center}
\end{subfigure}
\caption{Exact vs approximate expressions for $\Gamma_0$ (left) and
  $b(\Gamma_0-\Gamma_1)$ (right), which appear in the kinetic
  dispersion relation, Eq.~(\ref{flrdisp}). The approximations, given
  by Eqs. (\ref{gamma0approx}-\ref{gamma1approx}), are plotted for
  several choices of maximal Laguerre moment
  $\LL$. }
\label{fig:gammas}
\end{figure}

Here we show that the number of moments we keep affects the
order of accuracy of the FLR terms. This is because there will be a
maximum $\ell$ available to us directly in the equations. To illustrate
these issues, we will examine the linear dispersion relation of our
Laguerre-Hermite system in the artificial $k_\parallel=0$ slab limit
(with no collisions), which allows us to drop all terms in
Eq.~(\ref{glmevolve}) except the $i\omega_*$ drive
terms.  In this limit, the kinetic dispersion relation is given by
\begin{align}
\omega = \frac{\omega_{*,n}}{\tau^{-1}-\Gamma_0+1} \left[\Gamma_0 - \eta b (\Gamma_0 - \Gamma_1)\right], \label{flrdisp}
\end{align}
where $\omega_{*,n} \equiv \omega_* (a/L_n)$, and $\eta\equiv L_n/L_T$. Here $\Gamma_n(b) = I_n(b) e^{-b}$, where
$I_n(b) = i^{-n} J_n(ib)$ is the modified Bessel function.
The corresponding fluid dispersion for a system with maximum Laguerre moment $\LL$ is given by
\begin{align}
\omega = \frac{\omega_{*,n}}{\tau^{-1}-\Gamma_0+1}\sum_{\ell=0}^\LL \gm{}{\ell}
    \left[\gm{}{\ell}+\eta\left(\ell \gm{}{\ell-1}+2\ell\gm{}{\ell}+(\ell+1)\gm{}{\ell+1}\right)\right]. \label{fluid_disp}
\end{align}
Substituting the definition of $\mathcal{J}_\ell$ from Eq.~(\ref{jflr}) and taking the infinite moment limit $\LL\rightarrow\infty$, we obtain
\begin{align}
\omega&= \frac{\omega_{*,n}}{\tau^{-1}-\Gamma_0+1}\left[e^{-b}\sum_{\ell=0}^\infty \frac{1}{(\ell!)^2} \left(\frac{b}{2}\right)^{2\ell} - 
     \eta b e^{-b} \sum_{\ell=0}^\infty\left( \frac{1}{(\ell!)^2} \left(\frac{b}{2}\right)^{2\ell} 
   - \frac{1}{\ell!(\ell+1)!} \left(\frac{b}{2}\right)^{2\ell+1}\right) \right] \notag \\
&= \frac{\omega_{*,n}}{\tau^{-1}-\Gamma_0+1}\left[e^{-b} I_0 - \eta b e^{-b} (I_0 - I_1) \right] \notag \\
&= \frac{\omega_{*,n}}{\tau^{-1}-\Gamma_0+1}\left[\Gamma_0  - \eta b  (\Gamma_0 - \Gamma_1) \right]. \label{inf_disp}
\end{align}
Thus in the infinite moment limit, we obtain the correct dispersion relation. 
However, now consider the finite $\LL$ case (where we assert $\mathcal{J}_{\LL+1}=0$ for consistency): 
\begin{align}
\omega&= \frac{\omega_{*,n}}{\tau^{-1}-\Gamma_0+1}\Bigg[e^{-b}\sum_{\ell=0}^\LL \frac{1}{(\ell!)^2} \left(\frac{b}{2}\right)^{2\ell} 
    - \eta  b e^{-b} \sum_{\ell=0}^{\LL-1} \left(\frac{1}{(\ell!)^2} \left(\frac{b}{2}\right)^{2\ell} - \frac{1}{\ell!(\ell+1)!} \left(\frac{b}{2}\right)^{2\ell+1}\right)\Bigg]. 
\end{align}
Comparing term by term to \eqref{inf_disp}, we see that the finite $\LL$ case 
matches the kinetic dispersion relation only through order $4\LL$ in $k_\perp \rho$, since
\begin{align}
\Gamma_0 &= e^{-b} I_0 = e^{-b} \left[\sum_{\ell=0}^\LL \frac{1}{(\ell!)^2} \left(\frac{b}{2}\right)^{2\ell} \right] + \mathcal{O}(b^{2(\LL+1)}), \label{gamma0approx} \\
\eta b \Gamma_0 &= \eta b e^{-b} I_0=\eta b e^{-b} \left[\sum_{\ell=0}^{\LL-1} \frac{1}{(\ell!)^2} \left(\frac{b}{2}\right)^{2\ell} \right]+ \mathcal{O}(b^{2(\LL+1)}) ,\\
\eta b \Gamma_1 &=\eta b e^{-b} I_1 =\eta b e^{-b}\left[ \sum_{\ell=0}^{\LL-1} \frac{1}{\ell!(\ell+1)!} \left(\frac{b}{2}\right)^{2\ell+1} \right]+ \mathcal{O}(b^{2(\LL+1)}). \label{gamma1approx}
\end{align}
In Figure \ref{fig:gammas} we plot the result of the approximations above for several choices of $\LL$, along with the exact expression. As expected, the approximations are valid for larger $b$ as we increase $\LL$. Further, the approximations always eventually fall off to zero faster than the exact expressions. 

Dorland explored alternative FLR closures in his gyrofluid model in order to boost FLR accuracy \cite[]{D1},
since his model has $\LL=1$. Dorland settled on an FLR closure that corresponds to modifying Eq.~(\ref{jflr}) to
\begin{equation}
\hat{\mathcal{J}}_\ell = \frac{b^\ell}{\ell!} \frac{\partial}{\partial b} \Gamma_0^{1/2}(b), \label{eq:dorland_flr}
\end{equation}
which stems from approximating $\langle J_0 \rangle =\int d\mu B J_0 F_M \approx \Gamma_0^{1/2}$ instead of using the exact result, 
$\langle J_0 \rangle = e^{-b/2}$.  
This approximation is rigorous through second order in $k_\perp \rho$ while
imposing less gyroaveraging than the exponential for large $b$. Dorland then introduces
corresponding approximations in the calculation of the real space
density, $\bar{n}$, which are designed to exactly reproduce the
kinetic FLR behavior \cite[]{D1}. However, his procedure for
approximating $\bar{n}$ is not readily generalized to
arbitrary $\LL$.

\section{Laguerre-Hermite projection of magnetic drift terms} \label{app:toroidal}

We first write the gyrokinetic magnetic drift term as
\begin{equation}
{\bf v}_d \cdot\nabla h = (v_\parallel^2+\mu B) i\omega_{d} h 
  = \sum\limits_{\ell,m=0}^\infty 
    (v_\parallel^2+\mu B)\, \psi^\ell \phi^m \, i\omega_{d} \,
    H_{\ell, m}
\end{equation}
where (following \cite[]{BeerGF}) we have introduced the 
$i \omega_{d}$ operator, which combines the curvature and $\nabla B$
drifts into one expression, as is appropriate at low normalized plasma pressure:
\begin{equation}
i \omega_{d} \equiv (1/ B^2){\bhat}\times\nabla
B\cdot\nabla.
\end{equation}
The non-dimensionalized form of this operator has the species
dependencies in explicit form, $i \omega_{ds} = i \omega_d
(\tau_s/Z_s)$. Note that
${\bf v}_d\cdot\nabla= {\bf v}_d\cdot\nabla_\perp$, so we can use
the derivative relation from Eq.~(\ref{gradperp}) to move the
derivative operator inside the sum.  We next employ the recurrence
relations from Eqs.~(\ref{recurL}) and (\ref{recurHe}) to express this
as
\begin{align}
{\bf v}_d \cdot \nabla h  
  = &\sum\limits_{\ell,m=0}^\infty 
    \psi^\ell \, \left[\sqrt{(m+1)(m+2)} \phi^{m+2} +
    (2 m+1)\phi^m +\sqrt{m(m-1)}\phi^{m-2}\right]  i\omega^{(\kappa)}_{ds} H_{\ell,m} \notag \\
  +&\sum\limits_{\ell,m=0}^\infty 
     \phi^m \, \left[(\ell+1)\psi^{\ell+1} + (2\ell+1)\psi^\ell + \ell
     \psi^{\ell-1}\right]i\omega^{(\nabla B)}_{ds} H_{\ell,m}, 
\end{align}
where we have noted how one can distinguish the roles of the $\nabla
B$ drift $\propto \omega^{(\nabla B)}_{ds}$ and curvature drift
$\propto \omega^{(\kappa)}_{ds}$ if one wished to make that
distinction. To put this expression in the form of a Laguerre-Hermite transform, we
shift the indices:
\begin{align}
{\bf v}_d \cdot \nabla h
  = &\sum\limits_{\ell,m=0}^\infty \psi^\ell \phi^m 
   i \omega_{ds} \Big[ \sqrt{(m+1)(m+2)} \, H_{\ell,m+2}
   + (\ell+1) \, H_{\ell+1,m} \notag \\
   &\quad+ 2 \, (\ell+m+1) \, H_{\ell,m} 
   + \sqrt{m (m-1)} \, H_{\ell,m-2} + \ell \, H_{\ell-1,m} \Big].
\end{align}
The final step is then to use orthogonality to project this expression 
onto the Laguerre-Hermite basis, which is now trivial. 
We accomplish this by multiplying by projection functions and integrating over velocity. In non-dimensional
form, the result is
\begin{align}
 2\pi  \int_{-\infty}^\infty &d v_\parallel
   \int_0^\infty d\mu B\ \psi_\ell
   \phi_m\ {\bf v}_d \cdot\nabla h
   = 
   i \omega_{ds} \Big[ \sqrt{(m+1)(m+2)} \, H_{\ell,m+2}
   + (\ell+1) \, H_{\ell+1,m} \notag \\
   &+ 2 \, (\ell+m+1) \, H_{\ell,m} 
   + \sqrt{m (m-1)} \, H_{\ell,m-2} + \ell \, H_{\ell-1,m} \Big].
\label{projection}
\end{align}
This produces the magnetic drift terms for the $dG_{\ell, m}/dt$
equations. The $v_\parallel^2$ dependence of the curvature drift
couples $m$ moments to $m+2$ and $m-2$ moments, and the $\mu
B$ dependence of the $\nabla B$ drift couples $\ell$ moments to $\ell+1$ and
$\ell-1$ moments.

\section{Pseudo-spectral alternative to Laguerre convolutions}
\label{app:HLeval}

\subsection{A standard dealiased pseudo-spectral algorithm}
\label{app:nl}
The quadratic nonlinear terms in the gyrokinetic equation introduce
convolutions into any spectral representation of the equations. This
is familiar in flux-tube gyrokinetics. The
standard approach is to employ a dealiased pseudo-spectral
algorithm. To illustrate the approach, we consider a typical nonlinear
term,
\begin{equation}
{\bf v}_E \cdot \nabla n = {\partial \Phi \over \partial x} 
{\partial n \over \partial y} -  {\partial \Phi \over \partial y}
{\partial n \over \partial x},
\label{nlps}
\end{equation}
in which $x$ and $y$ are understood to be appropriately defined
perpendicular field line following coordinates. Given a perpendicular
coordinate-space Fourier representation of $\Phi$ and $n$ with $N$
modes in each direction, one first computes the derivatives as
multiplications by $i k_x$ and $i k_y$. Then one uses a discrete
Fourier transform to find a discretized representation of each
quantity on a grid with $3N/2$ grid points in each direction. The
fields are thus known on a mesh with $9/4$ more grid points than
spectral modes. The extra grid points are important in the dealiasing
step.

Eq.~(\ref{nlps}) is evaluated on the higher resolution spatial
grid by pointwise multiplication and addition. The result is then
transformed back to the lower resolution spectral representation with
an inverse discrete Fourier transform. Information contained in the
high-frequency (unresolved) modes is simply discarded. Surprisingly,
this ``dealiasing'' process recovers the exact spectral
convolution \cite[]{Orszag71}, including conservation of energy. Whether
or not one should dealias is a matter of debate, but the basic idea is
that dealiasing is more important as sharp derivatives appear in the
solution. For smoothly varying, random fields that often characterize
turbulence, dealiasing is frequently not done.

\subsection{Pseudo-spectral nonlinearity in the Laguerre-Hermite representation}
Relative to the example above, Eq.~(\ref{convective}) is complicated
by the convolution over Laguerre moments that is induced by finite
Larmor radius effects. There is no corresponding Hermite convolution,
but only because the FLR-averaged ${\bf E} \times {\bf B}$ drift
corresponds to $m = 0$ in Hermite space. (An electromagnetic
nonlinearity would have $m = 1$ Hermite structure and would require
a three term convolution.) At low resolution, the Laguerre convolution can be
carried out without much difficulty, using Eqs.~(\ref{convective}-\ref{Ckmn}). For more
than a few Laguerre moments, it will be more efficient to use the
analogue of the Fourier dealiased pseudo-spectral method. Our approach
is standard and well-described in \cite{Boyd}. 

We seek accurate integrals of the product of three Laguerre functions,
as is clear from Eq.~(\ref{Ckmn}). This implies that we must be able
to integrate polynomials of degree $3 \LL$ accurately. We will proceed by transforming to
a discretized $\mu B$ grid, evaluating the nonlinear term point-wise, and
transforming back. To obtain spectral accuracy and avoid aliasing, 
we pad the $H$ arrays with zeroes
in the Laguerre dimension, for $\ell = \LL+1$ to $J$. We need $2J+1
\ge 3 \LL$, or $J \ge (3 \LL - 1)/2$. Thus the ``three-halves'' rule
is replaced by this expression for $J$ in our problem. For $\LL=1$, as
in the Beer-Hammett equations, there is the surprise that we do not
need to dealias the Laguerre moments, but this is not true in
general. Note that because our indices start from zero, there are
$J+1$ moments and $J+1$ quadrature points, denoted by $x_j = (\mu
B)_j$. The specific values of the $x_j$ can be found from
$\psi_{J+1}(x_j) = 0$. The weights that are used to perform integrals
on this grid can be precalculated from
$$
  w_j = \int\limits_0^\infty dx \, \frac{\psi^{J+1}(x)}{\psi_{J+1}'(x_j) (x-x_j)}.
$$

To execute this algorithm, one would precalculate $\psi^\ell(x_j)$, 
$\psi_\ell(x_j)$, and $w_j$. These are the items required to
move back and forth from the discrete $\mu B$ grid to the moment
representation. For example, if we denote the distribution function
$h$ on the discrete $\mu B$ grid determined by the value of $J$ as
$h_J(k_x, k_y, z, m, x_j)$, then
$$h_J(k_x, k_y, z, m, x_j) = \sum\limits_{\ell=0}^\LL \psi^\ell(x_j)
H_{\ell,m}(k_x, k_y, z).$$ The sum does not need to go beyond $\LL$ for
the obvious reason that all of the $H_{\ell,m}$ moments are zero in
that range, but there are $J+1$ rows in this matrix, as required to
obtain the values of $h_J(x_j)$. The inverse transform requires the
$\psi_\ell(x_j)$ matrix, and the weights required for the integration,
$w_j$. It is convenient to store these as one matrix,
$T_{\ell j} = \psi_\ell(x_j) w_j$, so that
$$H_{\ell, m}(k_x, k_y, z) = \sum\limits_{j=0}^J T_{\ell j} h_J(k_x,
k_y, z, m, x_j).$$ In this direction, there are $J+1$ columns, but only
$\LL+1$ rows in the transformation matrix. 

This pseudo-spectral evaluation of the nonlinear term is exactly equivalent 
to the spectral convolution form in Eq. (\ref{convective}). 
The procedure will be expensive for large values of $\LL$, but not
more expensive than evaluating the convolutions
directly. For modest values of $\LL$, it is likely that the highly optimized nature of
matrix-vector multiplication on our target hardware (GPUs) will make
the pseudo-spectral approach inexpensive. 
For very small values of $\LL$ the pseudo-spectral
approach is not required: the convolutions
only give a few terms, which one can evaluate by hand as was done
in the Dorland, Beer, and Snyder gyrofluid models.

\section{Collision operator}
\label{app:collisions}

\subsection{The field-particle terms} \label{app:conservation}
The field-particle or integral components of $C(h)$ ensure conservation laws by construction.
The real-space moments which appear in field-particle terms are expressed in Laguerre-Hermite form in
Eqs.~(\ref{eq:ubarHL}-\ref{eq:tbarHL}). These scalar quantities, such as $\bar{u}_\parallel$,
determine {\it how much} of a particular quantity (such as real-space parallel momentum) needs to be
restored. The coefficients of these real-space moments in the second
line of Eq.~(\ref{eq:fullC}) determine the velocity-space structure of
the restoring terms. For example, the parallel momentum is restored with
$+ \nu \bar{u}_\parallel v_\parallel J_0 F_M$
which can be recast as 
$$
\nu \bar{u}_\parallel v_\parallel J_0 F_M = \nu \bar{u}_\parallel
\psi^0 \phi^1 J_0 = 
\nu \bar{u}_\parallel \psi^0 \phi^1 \left( \sum\limits_{k=0}^\infty
  \psi_k \mathcal{J}_k  \right).
$$
Projecting this term onto the Laguerre-Hermite basis as in Eq. (\ref{projection}), we have
\begin{align}
 2\pi  \int_{-\infty}^\infty d v_\parallel
   \int_0^\infty d\mu B\ \psi_\ell
   \phi_m\ \nu \bar{u}_\parallel \phi^1 \psi^0 \left( \sum\limits_{k=0}^\infty
  \psi_k \mathcal{J}_k  \right)
&= \nu \bar{u}_\parallel \delta_{m1} \sum_{k=0}^\infty \mathcal{J}_k \delta_{\ell k} = \nu \bar{u}_\parallel \delta_{m1} \mathcal{J}_\ell.
\end{align}
Here, the finite Larmor radius average that is expressed by the Bessel function
results in a convolution over the Laguerre basis, as described in Eq.~(\ref{Ckmn}). In the case of the
parallel momentum, one of the Laguerre basis functions in the triple product is $\psi^0$, so
the convolution reduces to a simple delta function,
$\delta_{\ell k}$. The dependence on $\phi^1$ produces the $\delta_{m1}$. So in
the end, the parallel momentum conservation term appears in the
$m=1$ equations, for all values of $\ell$ [see Eq.~(\ref{glmevolve})]. 
The rest of the conservation terms are entirely analogous.

The conserving terms in Eq.~(\ref{eq:fullC}) manifestly maintain
constant momentum and energy. The number of particles is also 
conserved, as is clear from the initial form of this collision
operator as a divergence in velocity space. Finite Larmor radius
effects obscure local number conservation, as it would appear that the
leading order term $- \nu b H_{0,0}$ represents local classical
diffusion. In gyrokinetics, this term is cancelled at long wavelength
by the term that enforces perpendicular momentum conservation,
$+ \nu \sqrt{b} \mathcal{J}_0 \bar{u}_\perp$. 

Parallel momentum conservation is manifest, but it is instructive to
consider the small-$b$ limit of Eq.~(\ref{glmevolve}) nonetheless,
assuming no finite-$\ell$ fluctuations (only for clarity):
$$
{d G_{0,1} \over dt} + \dots =- \nu (b+1) H_{0,1} 
+ \nu \, \mathcal{J}_0 \bar{u}_\parallel
= - \nu (b+1) H_{0,1} + \nu \, (1 - b/2) H_{0,1}
= -\nu {3 b \over 2} H_{0,1} + {\cal O}(b^2).
$$
In the long wavelength limit 
$(b\rightarrow 0)$, the guiding center parallel momentum is
conserved. Conservation of the real-space momentum is clear from the
role of $\bar{u}_\parallel$.

We have chosen the orthonormal form of the probabilists' Hermite
polynomials. This slightly obscures energy conservation, because of
the factor of $1/\sqrt{m!}$ in the $m=2$ equation. That is,
$T_\parallel = \sqrt{2} G_{0,2}$. To see that energy is conserved in
the long wavelength limit, we consider the effect of collisions on the total guiding
center temperature $T = {1 \over 3} (T_\parallel + 2 T_\perp)$ and
take the $b \rightarrow 0$ limit:
\begin{align*}
{d T \over dt} & = 
{1 \over 3} \left( \sqrt{2} { dG_{0,2} \over dt} + 2 {dG_{1,0} \over dt}
\right) + \dots \\ \notag
& = -{\nu \over 3} (b + 2) \sqrt{2} H_{0,2} + {2 \nu \over
  3} T - {2 \nu \over 3} (b + 2) H_{1,0} + {2 \nu \over 3} b (H_{0,0}
+ H_{1,0}) - {4 \nu \over 3} {b \over 2} T + {4 \nu \over 3} H_{1,0} \\ \notag
& \simeq -{2 \nu \over 3} T_\parallel + 2 \nu  T - {4 \nu \over
  3} T_\perp \\ \notag
& = 2 \nu \left[ T - {T_\parallel + 2 T_\perp \over 3} \right] = 0.
\end{align*}

\subsection{Collisions and free energy}

The collision operator will not increase the free energy as long as
the right hand side of Eq.~(\ref{Wcdot}) is non-positive. By
construction, $\partial W_C / \partial t = 0$ for any perturbation
that is Maxwellian in real space. In the previous section we
demonstrated conservation of number, parallel momentum and energy in
the $b \rightarrow 0$ limit. That is, 
\begin{align*}
 \frac{1}{\nu \tau_s} \pderiv{W_C}{t}|_{\rm fluid}
& \equiv - b |H_{0,0}|^2 - |H_{0,1}|^2 - 2 |H_{0,2}|^2  - 2
  |H_{1,0}|^2   + |\bar{u}_\parallel|^2 + |\bar{u}_\perp|^2 +
 3 |\bar{T}|^2  \\ 
& =  - b |H_{0,0}|^2 - |H_{0,1}|^2 - 2 |H_{0,2}|^2  - 2
  |H_{1,0}|^2   + |H_{0,1}|^2 + b |H_{0,0}|^2 
+ {1 \over 3} \left( T_\parallel + 2 T_\perp \right)^2 \\
& =  - 2 |H_{0,2}|^2  - 2 |H_{1,0}|^2   
+ {1 \over 3} \left| \sqrt{2} H_{0,2} + 2 H_{1,0} \right|^2 \\
& \leq 0. 
\end{align*}
The triangle inequality was used to obtain the final result. Equality
is obtained when the fluctuations in free energy are Maxwellian,
$$
T = {1 \over 3} (T_\parallel + 2 T_\perp) = {1 \over 3} 
(\sqrt{2} H_{0,2} + 2 H_{1,0}).
$$
Upon enforcing this constraint and eliminating $H_{1,0}$ from
$\dot{W}_C$ by setting $$H_{1,0} = {1 \over 2} (3T - \sqrt{2}
H_{0,2})$$
so that 
$$
|H_{1,0}|^2 = {1 \over 4} \left( 9 |T|^2 - 6 \sqrt{2} H_{0,2} T + 2 |H_{0,2}|^2 \right),
$$
one finds
\begin{equation}
 \frac{1}{\nu \tau_s} \pderiv{W_C}{t}|_{\rm Maxwellian} 
\equiv - 2 |H_{0,2}|^2  - 
 {1 \over 2} \left( 9 |T|^2 - 6 \sqrt{2} H_{0,2} T + 2 |H_{0,2}|^2 \right)
 + 3 |T|^2 = 0.
\end{equation}
As advertised, the right hand side vanishes when $\sqrt{2} H_{0,2} = T =
T_\parallel = T_\perp$. Fluctuations that are Maxwellian (with a
single temperature) are not affected by collisions.

The form of the $\mathcal{J}_\ell$ operator ensures that the FLR
corrections to real-space quantities such as $\bar{u}_\parallel$ are
strictly sub-dominant, as can be shown by rearranging the terms the
expression for $\dot{W}_C$, but the existence of an $H$-theorem is more
easily shown before gyroaveraging, as we demonstrate in the following section.

To summarize, we have demonstrated that long wavelength, real-space
fluctuations of density, momentum and temperature are unaffected by
collisions in our model. In fact, the conservation laws for real space
density, momentum and energy are correctly implemented at all
wavelengths. Small scale fluctuations of guiding center density,
momentum and energy are not conserved as a consequence of finite
Larmor radius averaging. This is the correct physical behavior. 

For all other (non-Maxwellian in real space) fluctuations, the free
energy of this gyrokinetic model is strictly damped by
collisions. Pitch-angle and energy scattering both contribute to this damping.

\subsection{The H-theorem}
We began with the Dougherty collision operator, which was shown in
Sec.~(\ref{sec:collisions}) to be
$$
 C(h) = \nu {\partial \over \partial {\bf v}} \cdot \left[ {T_0
    \over M} {\partial h \over \partial {\bf v}} + {\delta T \over
    M} {\partial F_M \over \partial {\bf v}} + {\bf v} \, h - \delta {\bf
    u} \, F_M \right],
$$
Note that in this section we work with dimensional, non-normalized quantities.
Here we have $n_0 \delta {\bf u} = \int d^3{\bf v} \, {\bf v} h$ and $n_0 \delta T = \int
d^3{\bf v} \, \left( m v^2/3 - T_0\right) h$. An equivalent form of $C$ is
\begin{equation}
C(h) = \nu {\partial \over \partial {\bf v}} \cdot \left[
{T_0 \over M} F_M {\partial \over \partial {\bf v}} \left(
{h \over F_M} - {M v^2 \over 2 T_0} {\delta T \over T_0} - {M\over
  T_0} {\bf v} \cdot \delta {\bf u}
\right)
\right].
\label{app:C2}
\end{equation}
The entropy production rate is
\begin{equation}
{dS \over dt} = - \int d^3{\bf v} {h \over F_M} C(h). 
\label{app:Sdot}
\end{equation}
By construction, we have 
$$
\int d^3{\bf v} \, {\bf v} C(h) = 0 \qquad {\rm and} \qquad \int d^3{\bf v} \, v^2 C(h) = 0,
$$
so we are free to modify Eq.~(\ref{app:Sdot}) as follows:
$$
{dS \over dt} = - \int d^3{\bf v} \left(
 {h \over F_M}  - {M v^2 \over 2 T_0} {\delta T \over T_0} - {M \over
   T_0} {\bf v} \cdot \delta {\bf u}
\right)
C(h). 
$$
Inserting $C(h)$ from Eq.~(\ref{app:C2}), 
we can then integrate by parts to find 
\begin{equation}
{dS \over dt} =  \nu {T_0 \over M} \int d^3{\bf v} \, F_M
\left| {\partial \over \partial {\bf v}} \left(
 {h \over F_M}  - {M v^2 \over 2 T_0} {\delta T \over T_0} - {M \over
   T_0} {\bf v} \cdot \delta {\bf u}
\right)
\right|^2.
\end{equation}
In this form, it can be seen that $\dot{S} \geq 0$. To obtain $\dot{S}
= 0$, it must be the case that
\begin{equation}
{\partial \over \partial {\bf v}} \left(
 {h \over F_M}  - {M v^2 \over 2 T_0} {\delta T \over T_0} - {M \over
   T_0} {\bf v} \cdot \delta {\bf u}
\right) = 0
\end{equation}
for all ${\bf v}$, which implies
$$
 {h \over F_M}  - {M v^2 \over 2 T_0} {\delta T \over T_0} - {M \over
   T_0} {\bf v} \cdot \delta {\bf u} = \alpha,
$$
where $\alpha$ is the integration constant. This can be rearranged to show
\begin{equation}
h = \alpha F_M + {\bf v} \cdot \delta {\bf u} F_M {M \over T_0} 
+ F_M {M v^2 \over 2 T_0} {\delta T \over T_0},
\end{equation}
from which one can immediately conclude that $\dot{S} = 0$ if and only
if $h$ is a perturbed Maxwellian. 

The Dougherty collision operator (together with our gyroaveraged
version) is also self-adjoint. A third equivalent form of
Eq.~(\ref{app:C2}) is 
\begin{equation}
C(h) = \nu {\partial \over \partial {\bf v}} \cdot \left[
{T_0 \over M} F_M {\partial \over \partial {\bf v}} \left(
{h \over F_M} \right) - {\delta T \over T_0} F_M {\bf v} - 
F_M \delta {\bf u}
\right].
\end{equation}
We consider the expression 
$$
\int d^3{\bf v} \, {h_a \over F_M} C(h_b)
$$
for any two distribution functions $h_a$ and $h_b$. One can write 
\begin{align*}
\int d^3{\bf v} \, {h_a \over F_M} C(h_b) & = \nu \int d^3{\bf v} \, {h_a \over f_M} 
{\partial \over \partial {\bf v}} \cdot \left[
{T_0 \over M} F_M {\partial \over \partial {\bf v}} \left(
{h_b \over F_M} \right) - {\delta T \over T_0} F_M {\bf v} - F_M \delta {\bf u}
\right] \\ 
& = 
\nu \int d^3{\bf v} \, {h_a \over F_M} {\partial \over \partial {\bf v}}
  \left[
{T_0 \over m} F_M {\partial \over \partial {\bf v}} \left( {h_b \over
  F_M} \right) \right] - 
\nu {\delta T \over T_0} \int d^3{\bf v} {h_a \over F_M} 
{\partial \over \partial {\bf v}} \cdot \left( F_M {\bf v} \right) 
\\
&\qquad\qquad- \nu \delta {\bf u} \cdot \left(
\int d^3{\bf v} \, {h_a \over F_M} {\partial F_M \over \partial {\bf v}}
\right) \\
&=
\nu \int d^3{\bf v} \, {h_a \over F_M} {\partial \over \partial {\bf v}}
  \left[
{T_0 \over M} F_M {\partial \over \partial {\bf v}} \left( {h_b \over
  F_M} \right) \right] - 
\nu {\delta T \over T_0} 
\int d^3{\bf v} h_a \left( 3 - {M v^2 \over T_0} \right)
\\
&\qquad\qquad+ \nu {M \over T_0} \delta {\bf u}  \cdot
\int d^3{\bf v} \, h_a {\bf v}.
\end{align*}
One can then integrate the first term by parts to find
\begin{align*}
\int d^3{\bf v} \, {h_a \over F_M} C(h_b) =
& - \nu {T_0 \over M} \int d^3{\bf v} \, F_M 
\left( {\partial \over \partial {\bf v}} {h_a \over F_M} \right) \cdot 
\left( {\partial \over \partial {\bf v}} {h_b \over F_M} \right) \\
& + {\nu \over 3 n_0} 
\left[ \int d^3{\bf v} \, h_b \left( {M v^2 \over T_0} - 3 \right) \right]
\left[ \int d^3{\bf v} \, h_a \left( {M v^2 \over T_0} - 3 \right) \right] \\
& + {\nu M \over n_0 T_0} 
\left( \int d^3{\bf v} \, h_b {\bf v} \right) \cdot \left( \int d^3{\bf v} \, h_a {\bf v} \right).
\end{align*}
This expression is completely symmetric with respect to $h_a$ and $h_b$,
which implies 
\begin{equation}
\int d^3{\bf v} \, {h_a \over F_M} C(h_b) = \int d^3{\bf v} \, {h_b \over F_M} C(h_a), 
\end{equation}
{\it i.e.,} the collision operator is self-adjoint. Neither
gyroaveraging nor projecting onto the Laguerre-Hermite basis alters
these conclusions.

\section{Reproducing the Beer 4+2 toroidal gyrofluid model} \label{appendix:beer_closures}
We can reproduce the six moment (4+2) toroidal gyrofluid equation set of Beer and Hammett \cite[]{BeerGF} by
first truncating all moments with $2\ell+m>3$. This effectively keeps all moments up to order
$ v^3 $. As described above, the evolution equations for the kept moments couple 
to higher unevolved moments via parallel convection, curvature and $\nabla B$ drifts, and mirror terms.
For the parallel convection terms, unevolved moments in the $G_{0,3}$ and
$G_{1,1}$ equations are closed using the Landau damping closures of
Hammett and Perkins \cite[]{HP,D1}, which corresponds to
\begin{align}
2\nabla_\parallel G_{0,4} &= \frac{\beta_\parallel}{\sqrt{3}}\nabla_\parallel G_{0,2} 
    +{\sqrt{2}} {D_\parallel} |k_\parallel| G_{0,3}, \\
\sqrt{2} \nabla_\parallel G_{1,2} &= \sqrt{2} D_\perp |k_\parallel| G_{1,1},
\end{align}
where $\beta_\parallel = (32-9\pi)/(3\pi-8)$, $D_\parallel=2\sqrt{\pi}/(3\pi-8)$, 
and $D_\perp=\sqrt{\pi}/2$ as in \cite[]{D1}. 
For the toroidal drift terms, unevolved moments in the $G_{0,2}$, $G_{1,0}$,
$G_{0,3}$ and $G_{1,1}$ equations are closed using the toroidal closures developed
by Beer \cite[]{BeerGF}, where closure coefficients were fit numerically so that the 
toroidal response function for the fluid equations matched the kinetic response. 
Beer's closure corresponds to
\begin{align}
&i\omega_d(\sqrt{12} G_{0,4}+G_{1,2}) = \sqrt{2}|\omega_d|\left(\nu_1\sqrt{2} G_{0,2} 
    + \nu_2 G_{1,0}\right), \label{nuprime1}\\
&i\omega_d(\sqrt{2} G_{1,2}+2G_{2,0}) = 2|\omega_d|\left(\nu_3 \sqrt{2}G_{0,2} 
    + \nu_4 G_{1,0}\right), \\
&i\omega_d(\sqrt{20}G_{0,5}+G_{1,3}) = \frac{1}{\sqrt{6}} |\omega_d| \left( \nu_5 G_{0,1} + \nu_6' \sqrt{6} 
G_{0,3} +\nu_7' G_{1,1} \right),\\
&i\omega_d(\sqrt{6}G_{1,3}+2 G_{2,1})  = |\omega_d| \left( \nu_8 G_{0,1} + \nu_9' \sqrt{6} 
G_{0,3} +\nu_{10}' G_{1,1} \right). \label{nuprime4}
\end{align}
where $\nu_1-\nu_{10}$ are the closure coefficients defined in \cite[]{BeerGF}, and
$\nu_6'=(\nu_{6r},\nu_{6i}-11)$, $\nu_7'=(\nu_{7r},\nu_{7i}-3)$, $\nu_9'=(\nu_{9r},\nu_{9i}-1)$, and
$\nu_{10}'=(\nu_{10r},\nu_{10i}-7)$. The modifications of these closure coefficients
are related to differences in the definitions of the higher moments here compared to Beer. 
This is discussed in \ref{nuprime}. The critical part is that in practice, these modified
coefficients give exactly the same toroidal terms as in Beer. For the mirror terms, unevolved moments 
in the $G_{0,3}$ and $G_{1,1}$ equations are closed as in Beer with `Maxwellian' closures since 
these terms introduce no new dissipative processes; this corresponds to
\begin{align}
G_{0,4}\nabla_\parallel\ln B = G_{1,2}\nabla_\parallel\ln B = G_{2,0}\nabla_\parallel\ln B = 0.
\end{align} 
For details about what is meant by a `Maxwellian' closure in our
Laguerre-Hermite context, we again defer to \ref{nuprime}. Finally, as
indicated in Appendix \ref{appendix:flr}, we make a few changes to the FLR
terms to be consistent with the FLR approximations of Dorland and
Beer. We change the definition of the Laguerre FLR operator
$\mathcal{J}_\ell \rightarrow \hat{\mathcal{J}}_\ell$, defined in
Eq.~(\ref{eq:dorland_flr}). We also change the definition of the
non-Boltzmann part of the particle space density, Eq.~(\ref{eq:nbar}), to match Beer's
approximation:
\begin{equation}
\hat{\bar{n}} = \frac{1}{1+b/2}G_{0,0} - \frac{b/2}{(1+b/2)^2}G_{1,0} + \Gamma_0 \Phi,
\end{equation}
where the $\Phi$ term results from our definition of the particle
space density as an integral involving $h$, not $g$.  Thus by
truncating our Laguerre-Hermite moments to keep only moments up to
order $ v^3 $ and choosing the above closures consistent with Beer's
closures, we can reproduce the Beer $4+2$ toroidal gyrofluid model.

\section{Fixing non-orthogonality of Beer's moment definitions and closures} \label{nuprime}

Our Laguerre-Hermite moment definitions are generally consistent with Beer's moment
definitions, as indicated by Eq.~(\ref{Beermoments}). However, a difference arises in the definition of the
higher, unresolved moments, due to the fact that our Laguerre-Hermite
moments are defined to be orthogonal while Beer's were not. For
example, Beer defined
\begin{equation}
r_{\parallel,\parallel} = \int d^3{\bf v}\ \nvpar^4\ \dg. \label{rpar}
\end{equation}
Inserting our distribution function expansion \eqref{dg} into this
integral, we find 
\begin{equation}
r_{\parallel,\parallel} = \sqrt{12}G_{0,4} + 3 G_{0,0} + 6\sqrt{2} G_{0,2} = \sqrt{12} G_{0,4} + 3n +
6T_\parallel = \sqrt{12}G_{0,4} + r_{\parallel,\parallel M}.
\end{equation} 
Here $r_{\parallel,\parallel M} = 3n+6T_\parallel$ is ``the Maxwellian part
of the $r_{\parallel,\parallel}$ moment'' in Beer's terminology, since
this expression can be calculated by taking $g$ to be purely Maxwellian
in the integral. Beer's full result for $r_{\parallel,\parallel}$ is thus 
\begin{equation}
r_{\parallel,\parallel} = 3n + 6 T_\parallel + \delta
r_{\parallel,\parallel};
\end{equation}
in this case, therefore, $\delta r_{\parallel,\parallel} = \sqrt{12}G_{0,4}$
is the quantity to be evolved or closed in some fashion.

Now we make a subtle change in reasoning. Instead of conceptualizing some
high moment in terms of Maxwellian and non-Maxwellian parts, the
$r_{\parallel,\parallel M}$ terms are better understood as the
components of $r_{\parallel,\parallel}$ that are \textit{not}
orthogonal to the other moments (namely $n$ and $T_\parallel$). While
this difference does not change the results for even moments such as $r$,
differences do appear for odd moments. Consider, for example, the $s_{\parallel,\parallel}$
moment,
\begin{equation}
s_{\parallel,\parallel} = \int d^3{\bf v}\ \nvpar^5\ \dg.
\end{equation}
Inserting Eq.~\eqref{dg} for $\dg$ and integrating gives
\begin{equation}
s_{\parallel,\parallel}=10 \sqrt{6} G_{0,3}+ \sqrt{120} G_{0,5} = 10 q_\parallel + \sqrt{120} G_{0,5},
\end{equation}
where $q_\parallel$ appears because $s_{\parallel,\parallel}$ is not orthogonal to 
the lower moments. However, upon inserting a Maxwellian for $\dg$, Beer
calculated $s_{\parallel,\parallel M}=0$, and thus missed the
non-orthogonal part of the $s$ moments. Proceeding similarly, one can
show that $s_{\parallel,\perp} = q_\parallel+3q_\perp + \sqrt{6}G_{1,3}$ and
$s_{\perp,\perp} = 4 q_\perp + G_{2,1}$ are also not orthogonal.

These $s$ moments appear in the toroidal drift terms where they must
be closed. Beer composed the toroidal drift closures with three parts:
a ``Maxwellian" piece proportional to $i \omega_d$; dissipative
corrections proportional to $|\omega_d|/\omega_d$; and reactive
corrections, proportional to $i \omega_d$. Here, $\omega_d \sim {\bf
  k}\cdot {\bf v}_d$, where ${\bf v}_d$ represents the curvature and
magnetic drifts. Our reinterpretation of gyrofluid moments as a
Laguerre-Hermite expansion suggests that the toroidal drift closures
for a given moment $\langle v^n \rangle$ should instead have two
parts: a non-orthogonal part and an orthogonal part.  The
non-orthogonal part does not depend on closure decisions at all;
instead it arises simply from the fact that $v^n$ is not generally
orthogonal to the Laguerre-Hermite polynomials.  The orthogonal part
can then be closed with both dissipative and reactive terms, as Beer
did, to reproduce the kinetic linear dispersion relation as well as
one can.

In our reinterpretation of the Beer closures, Beer's closure
coefficients for the reactive terms are shifted to account for
non-orthogonality:
\begin{align}
&i\omega_d(s_{\parallel,\parallel}+s_{\parallel,\perp}) = i\omega_d(11
  \qpar + 3\qprp) +|\omega_d| \left( \nu_5 \upar + \left(\nu_6-11i
  \frac{|\omega_d|}{\omega_d} \right) \qpar + \left(\nu_7-3i
  \frac{|\omega_d|}{\omega_d} \right) \qprp \right) \notag
  \\ &\qquad\qquad\ = i\omega_d(11 \qpar + 3\qprp) +
  |\omega_d| \left( \nu_5 \upar + \nu_6' \qpar +\nu_7' \qprp
  \right),\\ &i\omega_d(s_{\parallel,\perp}+s_{\perp,\perp}) =
  i\omega_d(\qpar + 7\qprp) +|\omega_d|\left( \nu_8 \upar +
  \left(\nu_9-i \frac{|\omega_d|}{\omega_d} \right) \qpar +
  \left(\nu_{10}-7i \frac{|\omega_d|}{\omega_d} \right) \qprp
  \right)\notag \\&\qquad\qquad\ \ = i\omega_d(\qpar +
  7\qprp)+|\omega_d| \left( \nu_8 \upar + \nu_9' \qpar +\nu_{10}'
  \qprp \right).
\end{align}
For convenience we have defined $\nu_6'=(\nu_{6r},\nu_{6i}-11)$,
$\nu_7'=(\nu_{7r},\nu_{7i}-3)$, $\nu_9'=(\nu_{9r},\nu_{9i}-1)$, and
$\nu_{10}'=(\nu_{10r},\nu_{10i}-7)$, where the un-primed $\nu$ coefficients
take the same values as in Beer. We can then
convert the Beer toroidal closures to our $\Gjk$ representation; this
is given in Eqs.~(\ref{nuprime1}-\ref{nuprime4}).
We stress that these changes are simply a reinterpretation, and that in practice
our expressions give exactly the same terms as in Beer. This preserves the
outstanding fit to the kinetic linear response function
that Beer achieved with these closures. 

Our reinterpretation has one significant benefit, noted earlier by \cite{Scott05}: in
the absence of closures ({\it i.e.,} if one simply truncates the fluid
hierarchy), and without driving and dissipation, one can
identify the conserved free energy. This was not possible in the form
that Beer conceptualized his closures. This improvement occurs solely
as a result of moving a few terms that Beer identified as reactive
closure corrections to the category of terms that should appear
independently of closure decisions.

\label{appendix:beer_free_energy} Our calculation of free
energy conservation in our Laguerre-Hermite model in Section
\ref{sec:free_energy} ensures that the Beer gyrofluid model conserves
free energy in the absence of driving and collisions when all closure
terms are dropped (after making the reinterpretations addressed in Appendix
\ref{nuprime}). However, one would like to be able to show that the
free energy does not increase when the closure terms are
included. This is complicated by the fact that the Beer closure scheme
includes both dissipative and non-dissipative (reactive) terms. While
the dissipative terms by nature can only decrease the free energy, the
non-dissipative terms are more troubling. Nonetheless, the dissipative
and non-dissipative terms collectively serve to mimic physical damping
from phase mixing, so one would expect that the aggregate contribution
of these terms is to decrease the free energy.  It is then enough to
show that the eigenvalues of the closed system are all damped in the
absence of driving and collisions.

\section{Diagnostic quantities of interest}

Here we document some diagnostic integrals in the Laguerre-Hermite basis.
\subsection{Heat flux}
The normalized radial heat flux for species $s$ is defined as
\begin{equation}
Q_s = \int d^3{\bf r}\int d^3{\bf v} \ \vE\cdot\nabla\Psi \ 
\frac{v_\parallel^2 + v_\perp^2}{2} \gyavgr{h}
= \int d^3{\bf R}\int d^3{\bf v}\ \gyavgR{\vE} \cdot\nabla\Psi
\frac{v_\parallel^2 + v_\perp^2}{2} h 
\end{equation}
with $\Psi$ here denoting the flux label and serving as the radial
coordinate, and $\langle \dots\rangle$ denoting a gyroaverage. Using the first form, and defining
\begin{gather}
\bar{p} = \frac{\bar{p}_\parallel + 2 \bar{p}_\perp}{3}= \int d^3{\bf v}\ \frac{v_\parallel^2 + v_\perp^2}{3} \gyavgr{h},
\end{gather}
one can write 
\begin{align}
Q_s = & \frac{3}{2} \int d^3{\bf r} \, \vE \cdot \nabla \Psi \ \bar{p}
        = \frac{3}{2} \int d^3{\bf r} \, \vE \cdot \nabla \Psi \
        \frac{\bar{p}_\parallel + 2 \bar{p}_\perp}{3} 
        \notag \\
 = & \int d^3{\bf r} \,  \vE \cdot \nabla \Psi  
\sum_{\ell=0}^\infty 
\left\{ 
  \left[\ell \mathcal{J}_{\ell-1}+ (2\ell+\frac{3}{2})\mathcal{J}_\ell
     + (\ell+1)\mathcal{J}_{\ell+1}\right] \ H_{\ell,0} + \frac{1}{\sqrt{2}}\mathcal{J}_\ell H_{\ell,2} 
\right\}.
\end{align}
Note that since $\vE \cdot \nabla \Phi = 0$, the convection of the
integral of $h$ at fixed ${\bf r}$ is equal to the convection of the
integral of $g$ at fixed ${\bf r}$. In physical units, the heat flux for species $s$ is $Q_{{\rm phys},
  s} = (n_0 v_t T)_{\rm ref} \rho_*^2 Q_s$.

\subsection{Particle flux}
Similarly, the normalized radial particle flux for species $s$ is defined as
\begin{equation}
\Gamma_s = \int d^3{\bf r}\int d^3{\bf v} \ \vE\cdot\nabla\Psi\  \gyavgr{h}
= \int d^3{\bf r}\ \vE\cdot\nabla\Psi\ \bar{n},
\end{equation}
with $\bar{n}$ defined as in Eq. (\ref{eq:nbar}).
The Laguerre-Hermite projection is
\begin{equation}
\Gamma_s = \int d^3{\bf r} \ \vE\cdot \nabla\Psi\ \sum_{\ell=0}^\infty
\mathcal{J}_\ell H_{\ell,0}.
\end{equation}
In physical units, the particle flux for species $s$ is $\Gamma_{{\rm
    phys}, s} = (n_0 v_t)_{\rm ref} \rho_*^2 \Gamma_s$.

\subsection{Turbulent energy exchange}
In our ordering, gyrokinetic fluctuations do not result in net heating
of the plasma, but it is possible for one species to absorb more energy
from the fluctuations than another. That is, turbulent fluctuations
can mediate exchanges of energy among species. The relevant diagnostic
of the normalized turbulent heating rate of species $s$ is
\begin{align}
\mathcal{H}_s &= \int d^3{\bf R} \int d^3{\bf v} \, \frac{\tau_s h_s}{F_{Ms}} C(h_s) \notag
  \\&= \int d^3{\bf R}\, \tau_s \left[ \sum_{\ell=0}^\LL \sum_{m=0}^\MM - \nu (b +
    2 \ell + m) |H_{\ell, m}|^2 \right]
  + \nu (|\bar{u}_\parallel|^2 + |\bar{u}_\perp|^2 + 3 |\bar{T}|^2).
\end{align}
In physical units, this heating rate is $\mathcal{H}_{{\rm phys},s} =
(\rho_*^2 v_{t, {\rm ref}} / a) \ \mathcal{H}_s$.
\pagebreak
\bibliography{bdbib}
\end{document}